\documentclass[%
 superscriptaddress,
 amsmath,amssymb,
 aps,
 prb,
 twocolumn,
 longbibliography 
]{revtex4-2}

\usepackage{graphicx}
\usepackage{dcolumn}
\usepackage{bm}
\usepackage{xcolor}
\usepackage{tikz}
\usepackage{xfrac}
\usepackage{blindtext}
\usepackage{times}
\usepackage{multirow}
\usepackage{chemformula}
\usepackage{braket}

\usepackage{color}
\usepackage[colorlinks=true, urlcolor=blue, linkcolor=blue, citecolor=blue, pdftex]{hyperref}

\usepackage[normalem]{ulem}

\definecolor{darkblue}{HTML}{004D6B}
\definecolor{darkred}{HTML}{8c1515}
\definecolor{darkgreen}{HTML}{006400}
\usepackage{hyperref}
\hypersetup{
	colorlinks=true,
	urlcolor=darkred,
	citecolor=darkblue,
	linkcolor=darkred,
	breaklinks
}

\usepackage[caption=false]{subfig}

\graphicspath{{figures/}}

\newcommand{\be}{\begin{equation}}
\newcommand{\ee}{\end{equation}}

\newcommand{\bea}{\begin{eqnarray}}
\newcommand{\eea}{\end{eqnarray}}
\newcommand{\beal}{\begin{align}}
\newcommand{\eeal}{\end{align}}

\newcommand{\goto}{\rightarrow}

\newcommand{\abs}[1]{\left| #1 \right|} 
\newcommand{\avg}[1]{\left< #1 \right>} 

\begin{document}

\title{Exploring Two-dimensional Coherent Spectroscopy with Exact Diagonalization:\\ Spinons and Confinement in 1D Quantum Magnets} 

\author{Yoshito Watanabe}
\affiliation{Institute for Theoretical Physics, University of Cologne, 50937 Cologne, Germany}

\author{Simon Trebst}
\affiliation{Institute for Theoretical Physics, University of Cologne, 50937 Cologne, Germany}

\author{Ciar\'an Hickey}
\affiliation{School of Physics, University College Dublin, Belfield, Dublin 4, Ireland}
\affiliation{Centre for Quantum Engineering, Science, and Technology, University College Dublin, Dublin 4, Ireland}

\begin{abstract}
Two-dimensional coherent spectroscopy (2DCS) with terahertz radiation offers a promising new avenue for the exploration of many-body phenomena in quantum magnets. This includes the potential diagnosis of fractionalized excitations, for which linear response often struggles due to the indistinguishability of a continuum of fractional excitations from that caused by disorders or impurities. However, the interpretation of the complex results produced by 2DCS remains a challenge, and a general prediction of the spectral characteristics of different types of excitations has not yet been established. 
In this paper, we develop a numerical approach based on exact diagonalization (ED) to push our understanding of 2DCS towards different scenarios. We first validate our approach by comparing numerical ED and exact analytical results for the spectroscopic signatures of spinons in one-dimensional transverse field Ising model 
and develop how to deal with the inherently small system sizes in ED calculations. 
Augmenting the model by a longitudinal field, we demonstrate significant changes to the 2DCS spectrum upon the field-induced spinon pair confinement,
which can be rationalized in our ED calculations and from a ``two-kink'' model (in the absence of integrability).
One advantage of our ED approach is its possible extension to finite temperatures, which we explore using thermally pure quantum states 
and demonstrate to change the intensity and spectroscopical patterns of 2DCS when going beyond the integrable model.
Our numerically exact results provide a benchmark for future experiments and theoretical studies relying on approximation methods, 
and pave the way for the exploration of fractionalized excitations in quantum magnets.
\end{abstract}

\maketitle

\section{Introduction}

Terahertz (THz) spectroscopy, in which a THz pulse incident on a system is used to excite and probe its dynamics, is widely employed in the study of quantum materials \cite{Kampfrath2011, Morris2014, Zhang2018, Bera2021}. In particular, the THz range aligns well with the typical excitation energies of quantum magnets. The technique can be understood within the framework of linear response, providing information akin to the dynamical structure factor, but restricted to zero momentum. In conventional magnets, this form of linear response typically reveals sharp excitation modes, such as magnons or triplons. However, in systems with fractionalized excitations, linear response reveals a broad continuum of excitations, reflecting the underlying fact that local operators create multiple (deconfined) excitations \cite{Han2012, Mourigal2013, Wen2019}. Consequently, such systems present a profound challenge in distinguishing between a continuum arising from fractionalized excitations and one resulting from, for example, thermal or impurity-induced disorder \cite{Banerjee2016, Zhu2017, Shen2016, Kimchi2018, Ma2018}. Furthermore, even in the absence of disorder, extracting precise information from the continuum is notoriously difficult. This complexity highlights the need for complementary methods or approaches to accurately characterize dynamics in quantum materials, particularly in distinguishing the nuances of its excitation spectra.

A recently developed method to gain deeper insights into the dynamics of unconventional magnets is to expand THz spectroscopy to include \emph{multiple} excitation pulses \cite{Lu2017, Zhang2023, Zhang2024}. If one employs two THz pulses to probe the system and, in addition, varies the timing between the two pulses, one can perform two-dimensional coherent spectroscopy (2DCS) -- a technique that, in the optical range, has already been extensively used in fields such as quantum chemistry, to probe the structure of complex molecules \cite{Khalil2003}, and semiconductor physics, to probe the dynamics of excitons \cite{Cundiff2012}. At its heart, such a multi-pulse approach allows one to extract the \emph{non-linear} response of the system by subtracting the single pulse responses from the multi-pulse response. Then, by performing a two-dimensional Fourier transform over the time arguments $\tau$, the time delay between the two pulses, and $t$, the measurement time after the second pulse, the frequency spectrum of these non-linear responses can be revealed. With THz sources, it has been argued that 2DCS has the potential to become one of the more potent tools to explore many-body phenomena in quantum magnets, such as in studying spin waves \cite{Fava2023} or, more enticingly, fractionalized quasiparticles such as spinons \cite{Wan2019, Li2021, Potts2023} or Majorana fermions \cite{Choi2020, Mcginley2022, Krupunov2023, Kazem2023, Qiang2023}.

2DCS, being a probe of higher-order dynamical correlation functions, inherently contains more information than in linear response. One of the key capabilities of non-linear response is the detection of interactions/statistics between quasiparticles within the same species \cite{Fava2020, Fava2023}, as well as between different modes \cite{Mcginley2022, Zhang2023, Zhang2024}. Furthermore, recent theoretical work \cite{Wan2019}, has shown that 2DCS can clearly identify one of the most elementary forms of fractionalized excitations -- spinon excitations in the (exactly solvable) one-dimensional transverse field Ising model (1D-TFIM). The ``spinon echo" signal, represented by a sharp anti-diagonal line along $\omega_t = -\omega_\tau$, arises from the interference of phases accumulated by spinon pairs created by consecutive THz pulses. Weak perturbations have been argued to introduce a finite lifetime to the spinon pairs, which in turn manifests as an energy-dependent broadening of the spinon echo signal, thereby revealing the lifetime of the individual spinon excitations \cite{Hart2023}. The appearance of a similar (sharp) anti-diagonal feature as a signature of fractionalized excitations has also been shown in the Kitaev honeycomb model \cite{Choi2020}, distinguishing it from broad continua due to more trivial effects. 

\begin{figure}[t]
    \centering
    \includegraphics[width=0.95\columnwidth]{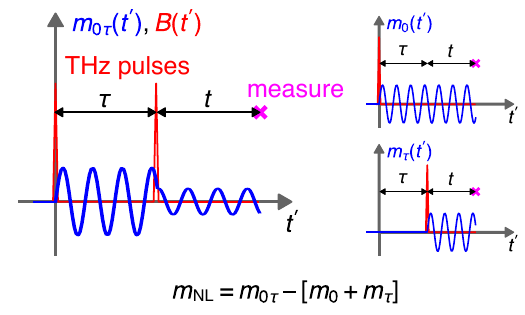}
    \caption{{\bf Schematic of the two-pulse measurement setup}. Two THz pulses (in red) are applied at times $t' = 0$ and $t' = \tau$, and the magnetization $m_{0\tau}$ (in blue) is measured at $t' = t+\tau$. Subtracting the single pulse responses, $m_0$ and $m_\tau$, from $m_{0\tau}$ isolates the non-linear magnetization $m_\text{NL}$.}
    \label{fig:tdcs_schematics}
\end{figure}

Despite the usefulness of 2DCS, theoretically, the calculation of the non-linear response functions and the interpretation of the 2DCS spectrum can be quite challenging. Here, we explore the utility of using exact diagonalization (ED) to study 2DCS in 1D quantum magnets. ED is well-suited to achieving long-time evolution, and correspondingly high-frequency resolution, but it is obviously rather limited in terms of system sizes. As such, it offers a different numerical trade-off than recent exploratory studies \cite{Sim2023, Gao2023} using a time-evolution of matrix product states (MPS), which can deal with much larger system sizes than ED but can be limited to shorter time scales due to the growth of entanglement under time evolution \cite{Schollwock2011}.
To showcase the applicability of our ED approach, we consider the 1D-TFIM and critically discuss finite-size effects in its exactly solvable limit and then extend our study to the case where the integrability is broken by a longitudinal field. 
We demonstrate that the fine resolution in the frequency domain achieved by ED reveals a detailed structure of the 2DCS spectrum characterized by a sequence of spinon-pair bound states. Finally, we will extend our approach to finite temperatures, a crucial developement for the comparison with experiments.

Our discussion in the following is structured as follows:
We start, in Sec.~\ref{sec:methods}, with a general overview of 2DCS, details on how we numerically compute the associated non-linear response functions with ED, and the model Hamiltonian that we focus on in this work. Next, in Sec.~\ref{sec:finite_size}, we investigate the origin of finite-size effects in the 1D-TFIM and discuss how to interpret the resulting 2DCS spectra. With the understanding of the finite-size effects in hand, we then turn to the case of a longitudinal field in Sec.~\ref{sec:breaking_integrability}. Combining our ED results with a ``two-kink'' approximation, we show the confinment of spinon pairs results in a significant change of the ``spinon-echo'' signal. Finally, in Sec.~\ref{sec:finite_temp}, we discuss the ED approach to finite temperatures and demonstrate how the finite-temperature effects can be captured in 2DCS. In Sec.~\ref{sec:discussion}, we summarize our results and discuss the experimental relevance of our findings.

\section{Methods}\label{sec:methods}

Before discussing any results, we first provide a brief overview of magnetic 2DCS and its theoretical underpinnings, details on how we numerically compute the associated non-linear response functions with ED, and the model Hamiltonian that we focus on in this work.  

\subsubsection*{Formalism of 2DCS}
We consider a simple time-dependent magnetic field consisting of two delta-function pulses at times $t'=0$ and $t'=\tau$. The magnetization is measured after the second pulse at a time $t'=t+\tau$. Thus, $\tau$ is the delay time between the two pulses and $t$ is the measurement time after the second pulse. The $\alpha$-component of the resulting time-dependent applied field can be written as $B^\alpha(t') = B^\alpha_0\delta(t') + B^\alpha_\tau\delta(t'-\tau)$.

We define $m^\alpha_{0\tau}(t,\tau)$ as the $\alpha$-component of the induced magnetization at time $t' = t+\tau$ after two successive pulses at $t' = 0$ with strength $B_0$ and at $t' = \tau$ with strength $B_\tau$. Analogously, $m^\alpha_{0}(t,\tau)$ and $m^\alpha_{\tau}(t,\tau)$ represent the induced magnetization after single pulses at $t' = 0$ with strength $B_0$ and at $t' = \tau$ with strength $B_\tau$ respectively. Subtracting these single-pulse magnetization responses ($m^\alpha_{0}$ and $m^\alpha_{\tau}$) from the two-pulse magnetization $m^\alpha_{0\tau}$ isolates the non-linear magnetization as
\begin{equation}
    \begin{split}
         m_{\text{NL}}^\alpha(t,\tau) =&\,\, m^\alpha_{0\tau}(t,\tau) - [m^\alpha_{0}(t,\tau) + m^\alpha_{\tau}(t,\tau)] \\
         =&\,\, \chi_{\alpha\beta\gamma}^{(2)}(t, \tau)B^\beta_\tau B_0^\gamma + \\ &\,\,\chi_{\alpha\beta\gamma\gamma}^{(3)}(t, \tau, 0)B_\tau^\beta B_0^\gamma B_0^\gamma +\\
        & \,\,\chi_{\alpha\beta\beta\gamma}^{(3)}(t, 0, \tau) B_\tau^\beta B_\tau^\beta B_0^\gamma + \dots,
    \end{split}
    \label{eq:M_NL}
\end{equation}
wherein only mixed cross-terms survive (for example, second-order terms proportional to $B_0^2$ and $B_\tau^2$ cancel out). This setup is schematically illustrated in Fig.~\ref{fig:tdcs_schematics}. Note that, in this two-pulse setup, there is a single second-order susceptibility but two distinct third-order contributions proportional to $B_\tau B_0^2$ and $B_\tau^2 B_0$ respectively. 

Concrete expressions for the non-linear susceptibilities can be derived using time-dependent perturbation theory \cite{mukamel1995}. In general, they are written as the equilibrium expectation value of nested commutators of the magnetization operators. As an example, the relevant third-order susceptibility can be neatly expressed in its most general form as 
\begin{equation}
     \chi^{(3)}_{\alpha\beta\gamma\delta}(t_3,t_2,t_1)=  \frac{2}{N} \textrm{Im}[R_1+R_2+R_3+R_4], \label{eq:R}
\end{equation}
where $t_1$, $t_2$ and $t_3$ are the time delays for a general multi-pulse setup, and the $R_a$ encode the contributions from the different possible orderings of the operators (we have suppressed their arguments and labels here for simplicity). As an example, $R_1$ is given by 
\begin{equation}
    R_1 = \avg{M^\gamma(t_1) M^\beta(t_2+t_1)M^\alpha(t_3+t_2+t_1) M^\delta(0)},
\end{equation}
where $M^\alpha(t) = \sum_i S_i^\alpha(t)$. By inserting resolutions of the identity, $R_1$ can be rewritten in terms of the energy eiegenstates $\ket{P}$, $\ket{Q}$ and $\ket{R}$ as 
\begin{equation}
    \begin{split}
        R_1 = \sum_{PQR} & m^\gamma_{0R}m^\beta_{RQ} m^\alpha_{QP} m^\delta_{P0} \\
        &\times e^{-\text{i}E_P t_1} e^{-\text{i}(E_P - E_R) t_2} e^{-\text{i}(E_P - E_Q)t_3},
    \end{split}
\end{equation}
where we have defined the magnetization matrix elements $m^\alpha_{fi}=\bra{f}M^\alpha\ket{i}$. See Appendix \ref{sec:2dcs} for the full expressions for $\chi^{(2)}$ and $\chi^{(3)}$ and their associated $R_a$. 

Setting $t_1 = 0, t_2 = \tau, t_3 = t$ and $t_1 = \tau, t_2 = 0, t_3 = t$ yields $\chi_{\alpha\beta\gamma\gamma}^{(3)}(t, \tau, 0)$ and $\chi_{\alpha\beta\beta\gamma}^{(3)}(t, 0, \tau)$ respectively. The two-dimensional Fourier transform of $\chi^{(3)}$ over positive $t$ and $\tau$ generates the 2DCS frequency spectrum. We define $\chi^{(3;1)}(\omega_t, \omega_\tau) \equiv \text{FT}[\theta(t)\theta(\tau)\chi^{(3)}(t, \tau, 0)]$ and $\chi^{(3;2)}(\omega_t, \omega_\tau) \equiv \text{FT}[\theta(t)\theta(\tau)\chi^{(3)}(t, 0, \tau)]$. It's important to note here that the positive time constraint, $t>0,\tau>0$, imposed by the form $\theta(t)\theta({\tau})\chi(t, \tau)$, results in a convolution of $[\delta(\omega_t) + 1/(\text{i}\pi\omega_t)][\delta(\omega_\tau) + 1/(\text{i}\pi\omega_\tau)]$ with $\tilde{\chi}(\omega_t, \omega_\tau)$, where $\tilde{\chi}(\omega_t, \omega_\tau)$ is the unconstrained Fourier transform of $\chi(t, \tau)$. Unfortunately, this convolution means that both the real and imaginary parts of the Fourier transform contain artificial broadening, e.g.~an additional $1/(\omega_t \omega_\tau)$ term for the real part. This distortion of the spectrum of the pure $\tilde{\chi}(\omega_t, \omega_\tau)$ is commonly referred to as ``phase twisting" \cite{Nandkishore2021, Hart2023}, and is a known impediment to the clean interpretation of 2DCS spectra \cite{Khalil2003, Kuehn2011}. Nevertheless, we focus here only on $\text{Re} [\chi^{(2)}(\omega_t,\omega_\tau)]$ and $\text{Im} [\chi^{(3;1,2)}(\omega_t,\omega_\tau)]$, the parts which, in the absence of the positive time constraint, would be non-vanishing.

\subsubsection*{Evaluation of non-linear susceptibilities with exact diagonalization}
We individually compute the non-linear susceptibilities $\chi^{(2)}$ and $\chi^{(3)}$ using the approach developed in Ref.~\cite{Gao2023}. Similar to the experimental setup, this involves applying two Dirac-$\delta$ pulses to the system and analyzing the resulting time evolution to determine the non-linear response. In the remainder, we focus on the case in which the field pulses and the measured magnetization are all aligned, so $\gamma=\beta=\alpha$. The action of a single pulse at time $t'$ on a state $\ket{\psi}$ is described by
\begin{equation}
    \ket{\psi'} = \exp\left(\text{i}B_{t'} M^\alpha\right)\ket{\psi},
\end{equation}
where $B_{t'}$ signifies the pulse magnitude. The wavefunction after two successive pulses, applied at times $t' = 0$ and $t' = \tau$ is
\begin{equation}
    \ket{\psi(t, \tau)} = e^{-\text{i}\mathcal{H}t}e^{\text{i}B_\tau M^\alpha}e^{-\text{i}\mathcal{H}\tau}e^{\text{i}B_0 M^\alpha}\ket{\psi}.
\end{equation}
The magnetization per site after two successive pulses is thus given by
\begin{equation}
    m^\alpha(t, \tau, B_\tau, B_0) = \frac{1}{N}\bra{\psi(t,\tau)}M^\alpha\ket{\psi(t,\tau)}.
\end{equation}
We can then calculate the individual non-linear susceptibilities as follows
\begin{equation}
    \begin{split}
        \chi^{(2)}_{\alpha\alpha\alpha}(t, \tau) & = \left.\frac{\partial^2 m^{\alpha}(t, \tau, B_\tau, B_0)}{\partial B_\tau\partial B_0}\right|_{B_0 = B_\tau = 0},\\
        \chi^{(3)}_{\alpha\alpha\alpha\alpha}(t, \tau, 0) & = \left.\frac{\partial^3 m^{\alpha}(t, \tau, B_\tau, B_0)}{\partial B_\tau \partial B_0^2}\right|_{B_0 = B_\tau = 0},\\
        \chi^{(3)}_{\alpha\alpha\alpha\alpha}(t, 0, \tau) & = \left.\frac{\partial^3 m^{\alpha}(t, \tau, B_\tau, B_0)}{\partial B_\tau^2 \partial B_0}\right|_{B_0 = B_\tau = 0}.
    \end{split}
\end{equation}
In practice, we numerically compute the derivatives using the central difference method. For example, $\chi^{(2)}_{\alpha\alpha\alpha}(t, \tau)$ can be computed using four different $(B_0, B_\tau)$ combinations. Setting $B_0=B_\tau = B$, it can be written as
\begin{equation}
    \begin{split}
        \chi^{(2)}_{\alpha\alpha\alpha}(t, \tau)  = \frac{1}{4 B^2}&\left[  m^\alpha(t, \tau, B, B) - m^\alpha(t, \tau, -B, B)\right.\\
        &\left.\!\!\!\! - m^\alpha(t, \tau, B, -B) + m^\alpha(t, \tau, -B, -B)\right].
    \end{split}
\end{equation}
This method is similarly applied to the calculation of the third-order derivatives for $\chi^{(3)}_{\alpha\alpha\alpha\alpha}(t, \tau, 0)$ and $\chi^{(3)}_{\alpha\alpha\alpha\alpha}(t, 0, \tau)$, using six combinations of $(B_0, B_\tau)$. Throughout, we use $B_0=B_\tau=0.001$. 

To make the Fourier transform well-behaved with finite time windows, a filter function $e^{-\eta(t^2 + \tau^2)}$ is applied, which broadens the signal in the frequency domain. We set a rather small broadening of $\eta = 0.001$, resulting in rather sharp peaks in the frequency domain. With this filter, we time evolve to a maximum time of $t_{\text{max}}=\tau_{\text{max}}=150$. The sampling intervals $\delta t$ and $\delta \tau$ determine the energy range as $\omega^{\text{max}}_t = \pi/\delta t$ and $\omega^{\text{max}}_\tau = \pi/\delta \tau$. We set $\delta t = \delta \tau = 0.25$.  
For the time-evolution and application of the pulses, both of which are achieved by exponential operator multiplications, we use the package ``Expokit.jl", a Julia implementation of EXPOKIT \cite{Sidje1998} that efficiently executes these calculations using Lanczos routines.

\begin{figure*}[tb]
    \centering
    \includegraphics[width=\linewidth]{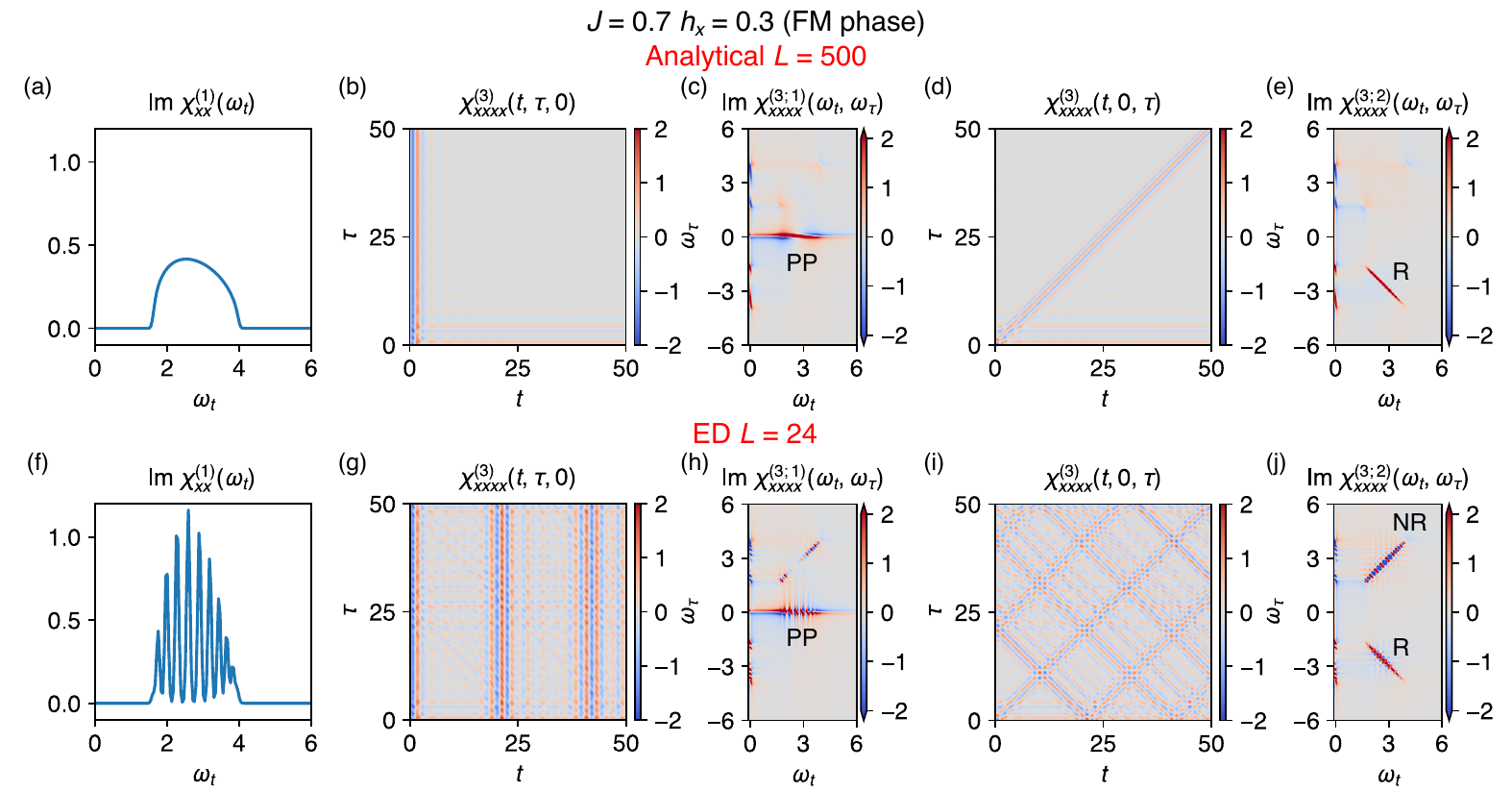}
    \caption{{\bf Revival of the signal in the small system size in one-dimensional transverse field Ising} (1D-TFIM). Two-dimensional coherent spectroscopy (2DCS) of 1D-TFIM with the model parameters $J=0.7$, $h_x = 0.3$, i.e., ferromagnetic phase. (a-e) Analytically calculated lnear-response $\text{Im}[\chi^{(1)}_{xx}(\omega_t)]$, third-order response $\chi^{(3)}_{xxxx}(t, t+\tau, t+\tau)$ and $\chi^{(3)}_{xxxx}(t, t, t+\tau)$ for $L = 500$, and 2d-FT of them. Spinon pairs manifest as continuous pump-probe (PP) signal at $\omega_\tau = 0$ in $\chi^{(3;1)}_{xxxx}$ and spinon-echo/rephasing (R) signal at $\omega_t = - \omega_\tau$ in $\chi^{(3;2)}_{xxxx}$. (f-j) The same quantities obtained by exact diagonalization (ED) for $L = 24$. Though the small $t$ and $\tau$ data are similar to the that of large system size, the signal starts to revive as $t$ and $\tau$ increase. The discrete character of the PP/R signal is evident. The non-rephasing (NR) signal at $\omega_t = \omega_\tau$ is also visible in $\chi^{(3;2)}_{xxxx}$, which is absent in the thermodynamic limit.}
    \label{fig:zero_temp_TFIM}
\end{figure*}

For the finite temperature simulations, we use the canonical thermal pure quantum state $\left|\phi\right> = \exp(-\beta H/2)\left|\phi_0\right>$ \cite{Sugiura2012, Sugiura2013}, where $\left|\phi_0\right>$ is a random vector whose norm is initialized to one, as the initial state. We average the results over $N = 10$ random initial states. See Appendix~\ref{sec:thermal_state} for the details of the finite temperature simulation.


\begin{figure*}[tb]
    \centering
    \includegraphics[width=0.8\linewidth]{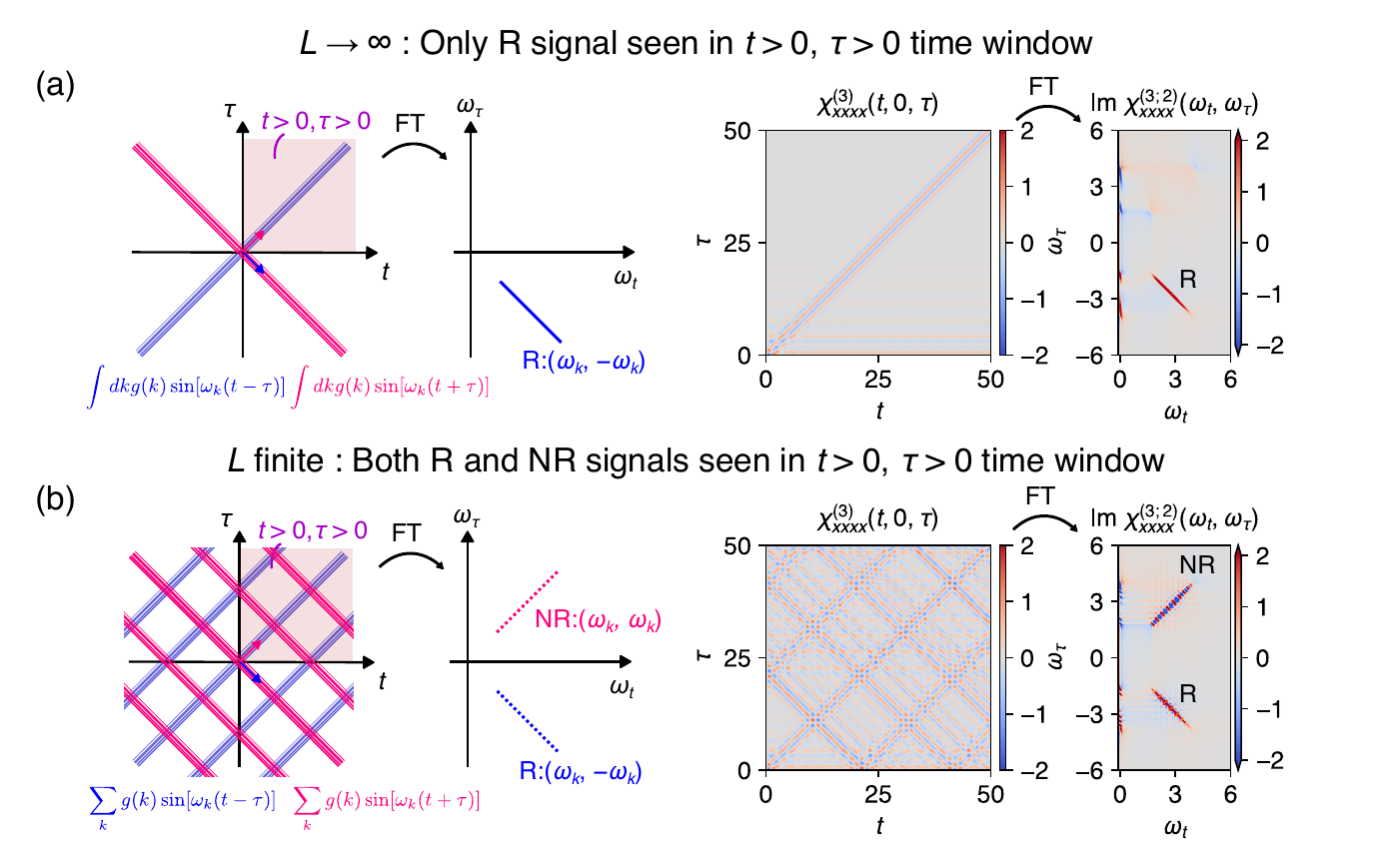}
    \caption{{\bf Schematics of the rephasing (R) and non-rephasing (NR) signal}. In the experiment, signals are measured at $t, \tau >0$ represented by the shaded area. (a) In the thermodynamic limit, the signal can be described by a superposition of continuous poles; therefore, it takes the form of the integral. R signal decays as $t + \tau$ increases, and NR signal decays as $t - \tau$ increases. FT of the signal for $t, \tau > 0$ only reveals R signal. (b) In the finite size system, the signal takes the form of the sum due to the discrete nature of possible $k$. R and NR signals are visible at $t, \tau > 0$ as a revival of the signal. FT of the signal yields discrete R and NR signals.}
    \label{fig:revival}
\end{figure*}

\subsubsection*{1D Transverse Field Ising Model (1DTFIM)}
The model we consider is the 1D-TFIM with the Hamiltonian
\begin{equation}
    \mathcal{H} = -J\sum_{i=1}^{L}\sigma_i^z\sigma_{i+1}^z - h_x\sum_{i=1}^L\sigma_i^x - h_z\sum_{i=1}^L\sigma_i^z,
\end{equation}
where $\sigma_i^{\alpha}$ $(\alpha = x, y, z)$ are Pauli matrices at site $i$, $J>0$ is the nearest neighbor ferromagnetic (FM) coupling, $h_x$ is the transverse field, and $h_z$ the longitudinal field. We assume periodic boundary conditions throughout. When $h_z = 0$, the model is exactly solvable via a Jordan-Wigner transformation, and, in the low-field ferromagnetically ordered phase, the elementary excitations are the kinks, or ``spinons", of the FM order (domain walls between blocks of aligned spins). These form a gapped dispersive band of excitations with energies $\lambda_k$. However, a single spin flip $\sigma_i^x$ generates a pair of spinons, with equal and opposite momenta. This form of fractionalization results in a continuum response at zero momentum when probed within the linear response regime.   

A finite longitudinal field $h_z$ breaks the integrability of the model, and generates a confining potential for the spinon pairs. Though the model is no longer exactly solvable, the resulting dynamics can be well described by a ``two-kink approximation", in which an effective Hamiltonian acting within the subspace of two spinon states is constructed \cite{Rutkevich2010, Coldea2010}. At large $h_z$, the spinons are tightly bound, and connect to the single spin-flip excitations of the high-field limit.

\section{Finite-size effects}
\label{sec:finite_size}
Since the available system sizes are limited in ED, we first need to have a solid understanding of potential finite-size effects. Here, we investigate finite size effects in two distinct scenarios, namely, (i) a magnetic field pulse creates a pair of excitations, meaning that linear response exhibits a continuum of excitations, and (ii) a magnetic field pulse creates only single excitations, meaning that linear response exhibits a discrete set of excitation modes. We will see that finite-size effects are more significant in the former scenario.

\subsection{Revival of signals}
If a system of interest has a continuous spectrum at zero momentum, then simulating 2DCS with a finite system size can lead to a spurious ``revival of signals" in the time domain. As a result, the Fourier-transformed signal in the frequency domain deviates from the true behavior expected in the thermodynamic limit. Here, we illustrate that the origin of this deviation is a combination of the discrete nature of the spectrum in finite-sized systems and the positive time constraint, $t>0,\tau>0$, enforced in the experimental setup.

In Ref.~\cite{Wan2019}, the full 2DCS spectrum for the exactly solvable 1D-TFIM was obtained (see Appendix~\ref{sec:tfim} for the derivation). The third-order susceptibility $\chi^{(3)}_{xxxx}$ reads:
\begin{equation}
    \begin{split}
        &\chi^{(3)}_{xxxx}(t_3, t_2, t_1) 
        = \frac{1}{L}\sum_{0<k<\pi} \left[A_k^{(1)} + A_k^{(2)} + A_k^{(3)} + A_k^{(4)}\right],
    \end{split}
\end{equation}
where
\begin{equation}
    \begin{split}
        A_k^{(1)} &= -8\sin^2\theta_k\cos^2\theta_k\sin(2\lambda_k (t_3 + t_2 + t_1)), \\
        A_k^{(2)} &= 8\sin^2\theta_k\cos^2\theta_k\sin(2\lambda_k (t_2 + t_1)), \\
        A_k^{(3)} &= -4\sin^4\theta_k\sin(2\lambda_k (t_3 + t_1)), \\
        A_k^{(4)} &= -4\sin^4\theta_k\sin(2\lambda_k (t_3 - t_1)),
    \end{split}
    \label{eq:chi3_tfim}
\end{equation}
with $\lambda_k$ the single-spinon energies, and the angle $\theta_k$ defined by $\tan\theta_k = (2J\sin k)/(2J\cos k -2h_x)$. 

Figure~\ref{fig:zero_temp_TFIM} compares the analytical and numerical $\chi^{(1)}_{xx}$ and $\chi^{(3;1,2)}_{xxxx}$ of the 1D-TFIM in the FM phase. The analytical data is obtained for a large system size of $L=500$ using Eq.~(\ref{eq:chi3_tfim}), representing the expected behavior in the thermodynamic limit, and the numerical data is obtained for $L=24$ with ED. For the ED data, we checked that the error between the analytical result and the ED result for the same system size is less than $10^{-5}$ in the time domain data, and no apparent increasing trend is evident during the chosen time evolution window.

There are unsurprisingly a number of similarities between the two system sizes. In $\chi^{(3;1)}_{xxxx}$, there is a pump-probe (PP) signal along the $\omega_t$ axis at $(\omega_t, \omega_\tau) =  (2\lambda_{k}, 0)$. In $\chi^{(3;2)}_{xxxx}$, there is an anti-diagonal signal at $\omega_t = -\omega_\tau = 2\lambda_{k}$ corresponding to the ``spinon-echo"  or rephasing (R) signal, originating from $A_k^{(4)}$. For the smaller system size, the discrete nature of these signals simply results from the discreteness of the momentum $k$. This is most evident in the linear response $\chi^{(1)}_{xx}$, where the $L=24$ system contains a visibly discrete set of $L/2$ peaks, corresponding to the $L/2$ allowed pairs of spinons with momenta $k$ and $-k$, while the larger system size effectively has a continuum of excitations due to its much larger number of $k$ points.   

There are two obvious differences though between the two system sizes. In the time domain, there are periodic revivals of the signal in the small system size used for the ED that do not appear in the larger system size, and, in the frequency domain, there are additional diagonal non-rephasing (NR) signals in the small system size which are absent in the larger system size. 
To understand this, let's first consider a one-dimensional time domain, and for simplicity assume that the energies of the spinons are equally spaced with a uniform frequency spacing of $\Delta\omega\propto 1/L$. If one superposes multiple sine waves, $\sum_{k>0} \sin(\omega_k t')$, with a uniform frequency spacing of $\Delta\omega$, then the wavelength of the resulting beating pattern will be proportional to $1/\Delta\omega$. In the thermodynamic limit, $L\goto \infty$, $\Delta\omega\goto 0$, and the spinon spectrum becomes continuous, which means that the wavelength diverges and the periodic signal disappears from the positive time axis, leaving a maxima at $t'=0$. 

In the two-dimensional time domain of relevance here, a similar scenario plays out. As illustrated in Fig.~\ref{fig:revival}, for $L\goto\infty$, we need to superpose a continuum of sine waves, $\int dk \, g(k) \sin[\omega_k (t\pm \tau)]$, with $g(k)$ simply representing matrix element factors. For the $t-\tau$ case, there is a maxima along the line $t=\tau$, and for the $t+\tau$ case, the maxima is along $t=-\tau$. Focusing on the relevant quadrant with $t>0,\tau>0$, only the $t=\tau$ line can be observed which, after Fourier transforming, is exactly the rephasing R signal. The $t=-\tau$ line is absent from the purely positive time window, and thus the absence of the corresponding non-rephasing NR signal is generally expected when the system has a continuous spectrum in the thermodynamic limit. On the other hand, for $L$ finite, we are adding a finite number of frequencies, $\sum_k \,g(k) \sin[\omega_k (t\pm \tau)]$, resulting in periodic signals with wavelengths roughly proportional to $L$. This means that, in the positive time quadrant, $t>0,\tau>0$, both rephasing and non-rephasing lines are present, and thus the Fourier transformed spectra have both signals present. The appearance of the NR signal is thus a consequence of the discreteness of the spectrum. In the ED, our time window for the FT in both $t$ and $\tau$ extends to several times $L$, so we indeed observe both R and NR signals and, similarly, we are able to resolve the discrete nature of the signals. 

It should be noted that there is a straightforward solution in this particular case to avoid such revivals of signals in the time domain. One can simply choose the time window small enough such that no revivals occur within it (alternatively, one can choose a large damping factor $\eta$ in the filter function $e^{-\eta(t^2+\tau^2)}$). Thus, the NR signal will be absent and the R signal will be continuous, both as expected for the thermodynamic limit. For large system sizes, this is relatively straightforward as the periodic revivals are anyway spread far apart in time, but for the small system sizes accessible with ED this heavily constrains the available time window and hence leads to an extremely poor resolution in frequency space. More importantly, it's, in general, not possible to know a priori whether the revivals are an artifact, to be avoided, or a feature, to be included, i.e.~whether the true physics is really a continuum or simply a dense set of discrete excitations (as, for example, observed in the confining scenario of the next section).   

In Fig.~\ref{fig:revival}(a), it is the positive time constraint, $t>0,\tau>0$, that ensures that it is only the R signal that is present in the thermodynamic limit. If we were to include both positive and negative $\tau$, then both the NR and R lines would be visible in the time domain, and hence one would observe both diagonal NR and anti-diagonal R signals in the 2DCS spectrum. In addition, the artificial broadening induced by the positive time constraint in the Fourier transform would be absent (in the case of the imaginary part). Taken together, this means that, for the relaxed constraint, $t>0$ only, the small system size ED result is simply a discretized version of the results in the thermodynamic limit, with all of the same qualitative features and signals (see Appendix \ref{sec:posneg_tau} for an illustration of this). 


\begin{figure}[tb]
    \centering
    \includegraphics[width=0.9\columnwidth]{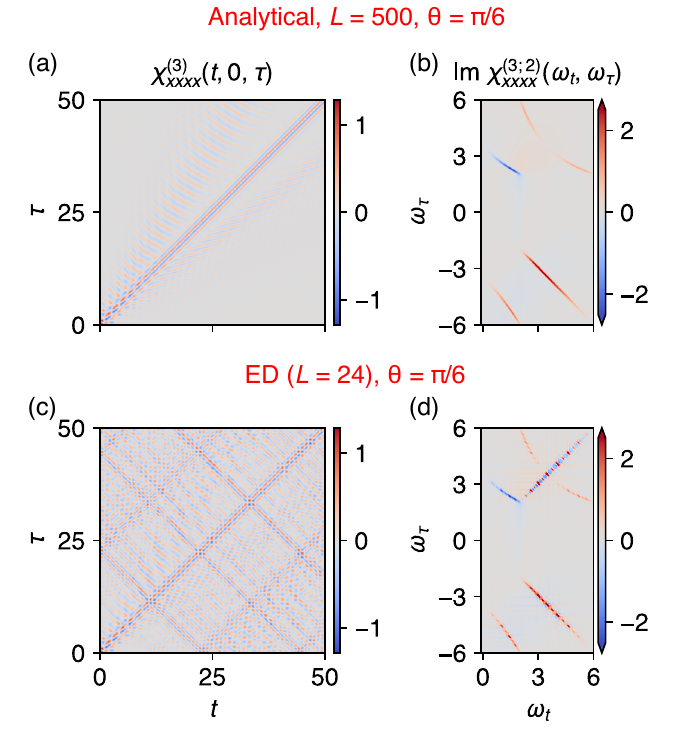}
    \caption{{\bf Twisted Kitaev Model}. (a) Analytical result for $L=500$. In addition to the R signal, there are streak features in both first and fourth quadrant, originating from the interplay between different spinon modes.
    (b) ED result for $L=24$. Additional diagonal NR signal at $\omega_t = \omega_\tau$ is observed.}
    \label{fig:TKM}
\end{figure}

\subsubsection*{Twisted Kitaev model}

As another example in which identical finite size effects are present, we show results for another exactly solvable model, the twisted Kitaev model (TKM) \cite{Zhang2015, You2016, Morris2021, Sim2023_2}. The Hamiltonian is
\begin{equation}
    \mathcal{H} = -J\sum_{i}^{L'}\left[\tilde{\sigma}_{2i-1}(\theta)\tilde{\sigma}_{2i}(\theta) + \tilde{\sigma}_{2i}(-\theta)\tilde{\sigma}_{2i+1}(-\theta)\right],
\end{equation}
where $L' = L/2$, $\tilde{\sigma}_i(\theta) = \sigma_i^z\cos(\theta/2) + \sigma_i^y\sin(\theta/2)$, and $\theta$ is the ``twist" angle. For $0 \leq \theta < \pi/4$, the ground state is a doubly degenerate FM state, polarized along the $z$-direction. The elementary excitations are again spinons (kinks, or domain walls, of the FM state), but now there are two kinds of spinons, with dispersions $l_k$ and $\lambda_k$, where $k = 2\pi n/L'$ and $n = 1, 3, \cdots, L'$ \cite{You2016}. 

Analytical expressions for the second and third-order susceptibilities for the TKM have been derived in Ref.~\cite{Sim2023_2}, with the result for $\chi^{(3;2)}_{xxxx}$ shown in Fig.~\ref{fig:TKM}(a) for a large system size of $L = 500$, again representing the behavior of the thermodynamic limit. Similar to the 1D-TFIM, a magnetic pulse along the $x$-direction only excites spinon pairs, with equal and opposite momenta; therefore there appears again a spinon-echo R signal. In addition, due to the interplay between the $l_k$ spinons and $\lambda_k$ spinons, there are additional streak features for both positive and negative $\omega_\tau$ \cite{You2016}

In ED, on an $L=24$ chain, we observe all of the features of the analytical result in the thermodynamic limit, but with discretized streaks due to the discrete number of $k$ points. Again, as in the TFIM, the main difference, in the frequency domain, is the additional diagonal NR signal, which is absent in the thermodynamic limit. The small system sizes available in ED are again able to capture the key qualitative features of the 2DCS spectrum, with the caveat that an additional NR signal presents itself.

\begin{figure}[tb]
    \centering
    \includegraphics[width=0.9\columnwidth]{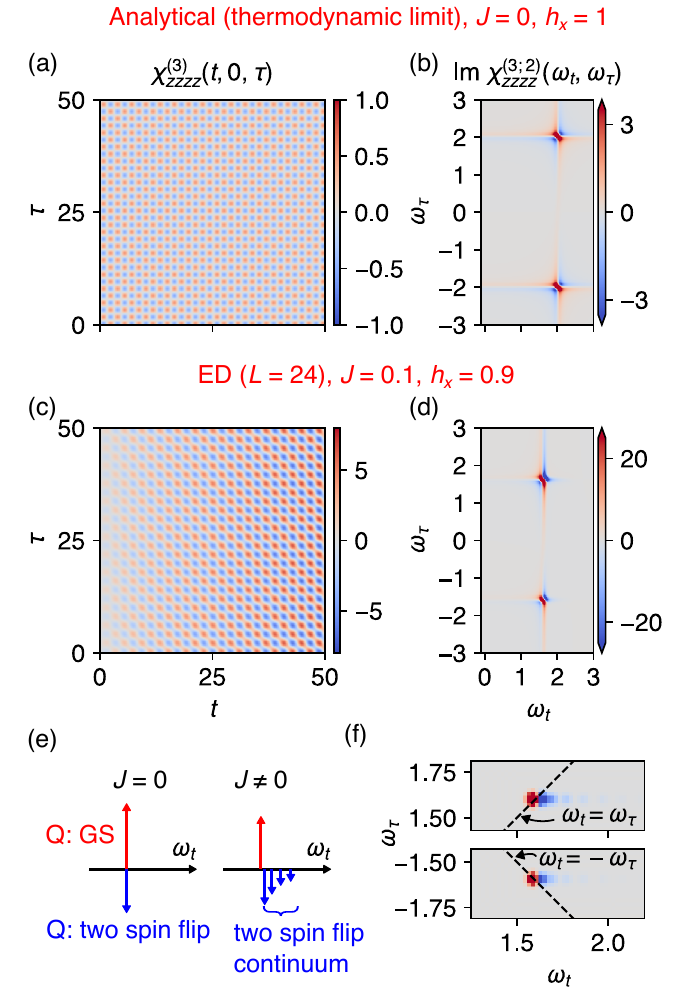}
    \caption{{\bf Single spin flip excitations in paramagnetic phase (PM)}. (a and b) The result for $J=0$, where the system size independent exact result is available. R and NR are observed at $(\omega_t, \omega_\tau) = (2h_x, -2h_x)$ and $(2h_x, 2h_x)$, respectively. (c and d) ED result for $L=24$ with $J=0.1$. (e) Schematics of the processes responsible for the NR signal. The intermediate state Q is either the ground states (red) or two-spin flip state (blue). In the presence of the interaction $J$, the two-spin flip states with zero total momentum constitute a continuum, which ends up with discrete poles in the finite size system. (f) Actual position of the poles for $L=24$. The data is obtained by Fourier transforming both positive and negative time data, which eliminates the phase twisting effect. Slightly different frequencies of the signals with opposite signs are superposed, which manifests diverging behavior in the time domain data (c).}
    \label{fig:PM}
\end{figure}

\begin{figure*}[tb]
    \centering
    \includegraphics[width=\linewidth]{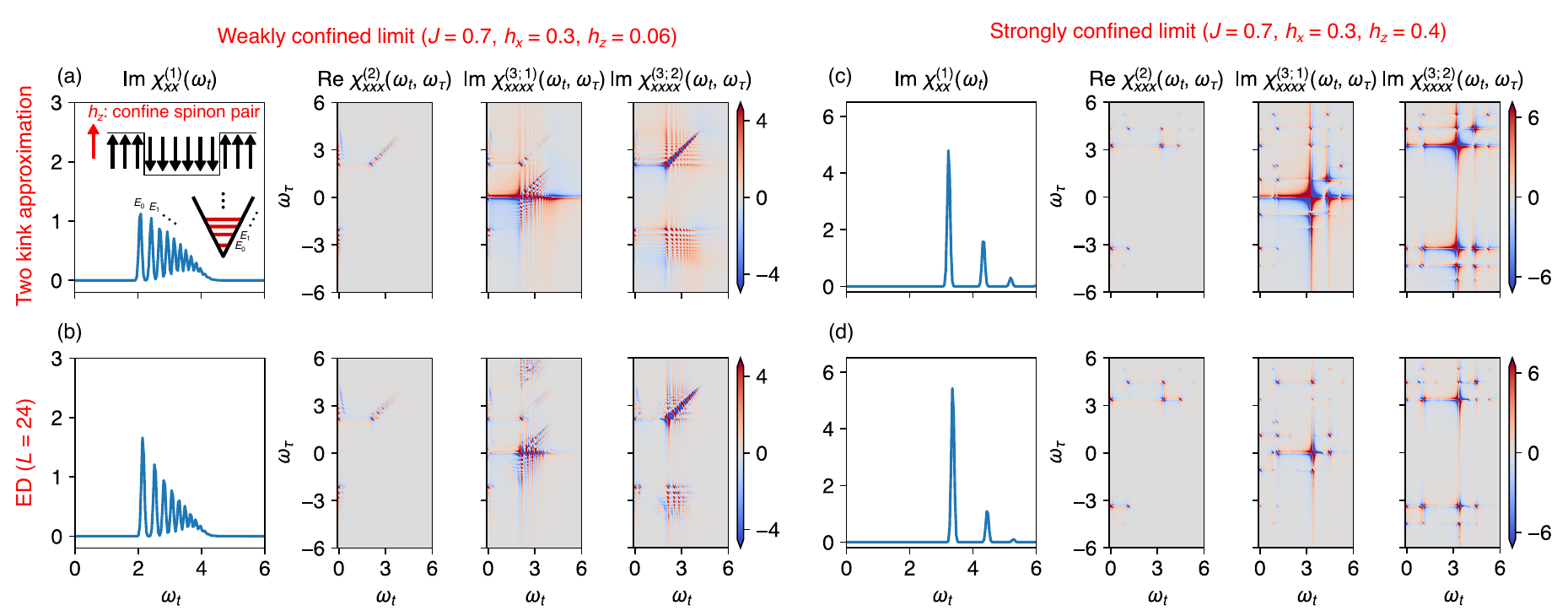}
    \caption{{\bf Spinon-confinement via an applied longitudinal field} with parameters $J=0.7, h_x = 0.06$ in (a, b) the weakly confined limit, $h_z = 0.06$, and (c, d) the strongly confined limit, $h_z = 0.4$. All relevant susceptibilities, $\chi^{(1)}_{xx}, \chi^{(2)}_{xxx}$, and $\chi^{(3;1,2)}_{xxxx}$ are shown. The results in (a), (c) (upper row) are obtained analytically using the two-kink approximation in the thermodynamic limit, while the results in (b), (d) (lower row) are obtained with ED for $L = 24$.}
    \label{fig:zero_temp_lfield}
\end{figure*}

\begin{figure*}[tb]
    \centering
    \includegraphics[width=0.9\linewidth]{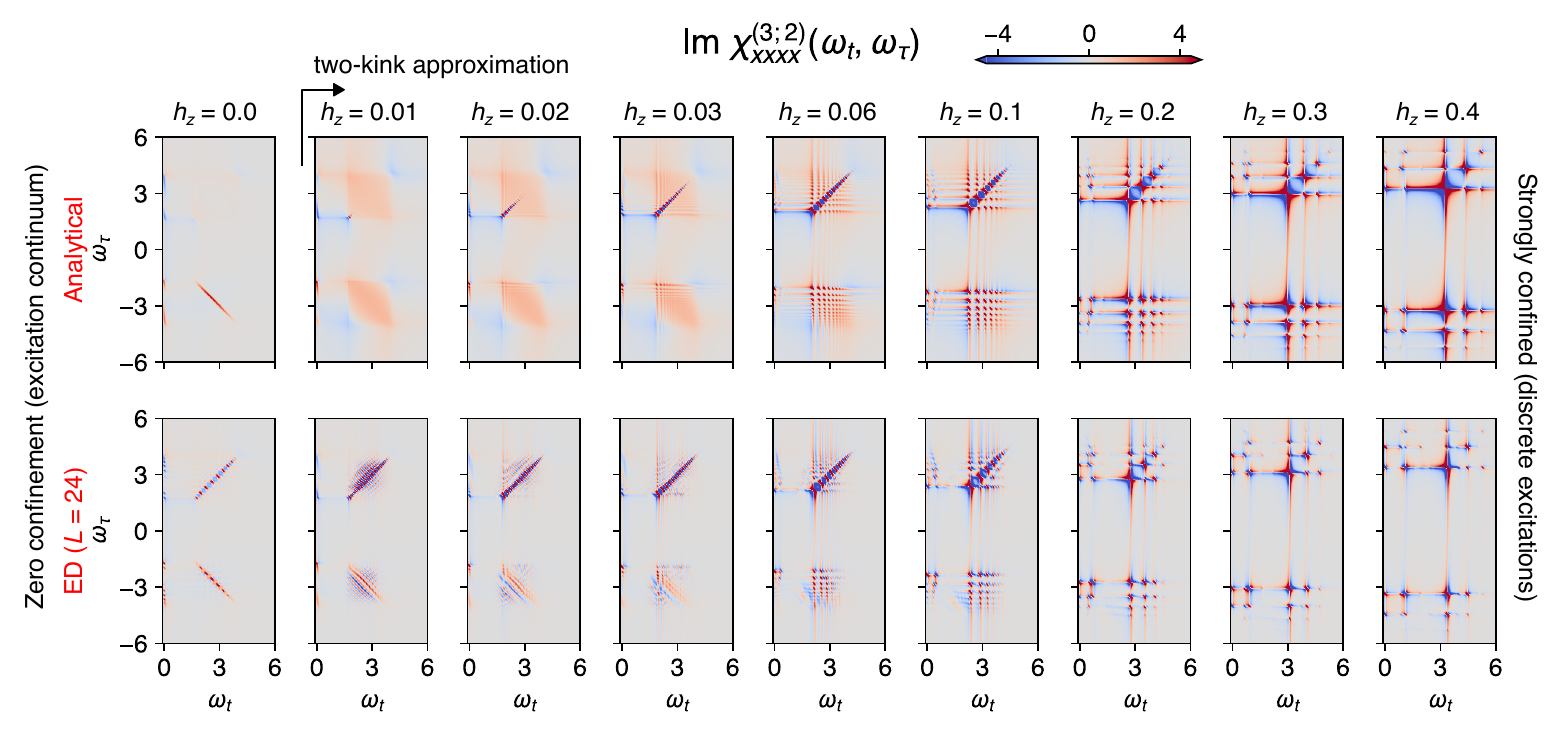}
    \caption{{\bf Evolution of the spinon-echo rephasing signal}. A comparison of $\chi^{(3;2)}_{xxxx}$, which contains the spinon-echo signal, for different longitudinal fields computed with the two-kink approximation (top row) and ED (bottom row). In both cases, the sharp spinon-echo signal at $h_z = 0$ is broadened as $h_z$ increases, ending as a set of discrete peaks in the strongly confined regime. Note that since the two-kink approximation does not include four-kink states, it overestimates the broadening of the spinon-echo signal.}
    \label{fig:lfield_detal}
\end{figure*}

\subsection{Discrete spin-flip excitations in the paramagnetic phase}
\label{sec:PM}

Next, we discuss finite size effects in the paramagnetic (PM) phase of the 1D-TFIM, i.e.~$h_x > J$. In this phase, a magnetic field pulse along the $z$-direction creates a single spin flip excitation with momentum $k = 0$, and the spectrum is expected to be discrete even in the thermodynamic limit. We thus expect that ED should be able to produce qualitatively similar results as those expected in the thermodynamic limit. Also, unlike the case of the FM ground state, we expect to observe an NR signal corresponding to single spin-flip excitations.

Figure~\ref{fig:PM}(a) and (b) show $\chi^{(3;2)}$ for the trivial limit of $J = 0, h_x =1$ ($h_x/J\goto\infty$), in which the result can be obtained analytically. In this transverse field-only limit, the ground state can be written as $\ket{0} = \ket{\rightarrow\rightarrow\cdots\rightarrow}$. Excitations involving either a single or double spin-flip at sites $i$ or $(i, j)$ are represented by $\ket{\leftarrow}_i = \sigma_i^z\ket{0}$ for a single site, and $\ket{\leftarrow}_{i}\ket{\leftarrow}_j = \sigma_i^z\sigma_j^z\ket{0}\, (i\neq j)$ for two sites, respectively. These are energy eigenstates with energies $E = 2h_x$ and $E = 4h_x$ respectively.

Keeping in mind that the matrix elements involved for $\chi^{(3)}$ can be written as $m^z_{0R}m^z_{RQ} m^z_{QP} m^z_{P0}$, we can think of two types of processes: one with $\ket{Q} = \ket{0}$, simplifying the matrix element to $|m^z_{0R}|^2 |m^z_{0P}|^2$, and another with $\ket{Q}$ involving double spin-flip states. In both cases, $\ket{P}$ and $\ket{R}$ are single spin-flip states. Noting that $(\sigma_z^i)^2$ is the identity operator, the number of contributions scales as $L^2$ for $\ket{Q} = \ket{0}$ processes, and as $2L(L-1)$ for processes with $\ket{Q}$ involving double spin-flip states. Combining these together, it can be shown that the summation of the different $R_a$ in Eq.~(\ref{eq:R}) for both types of processes results in a cancellation of the terms proportional to $L^2$, leaving only the terms proportional to $L$. Dividing by $L$, we thus obtain system size-independent peaks at $(\omega_t, \omega_\tau) = (2h_x, 2h_x)$ and $(2h_x, -2h_x)$, corresponding to NR and R signals respectively. Unlike in the low-field FM phase, the NR signal does not vanish here as the pole is now discrete.

Next, we consider the effects of a finite Ising interaction $J$, by which the spin-flip excitation obtains a dispersion. Figure~\ref{fig:PM}(c) and (d) shows the ED result with $L = 24$ for $J = 0.1, h_x = 0.9$. Similar to the non-interacting case, we observe NR and R signals at $(\omega_t, \omega_\tau) \approx (2h_x - 2J, 2h_x - 2J)$ and $(2h_x - 2J, -2h_x +2J)$, respectively, whose energies corresponds to that of the single spin-flip excitation at $k=0$. In the time domain data, the amplitude of the signal increases as $t$ increases, see Fig. \ref{fig:PM}(c). Similar to the non-interacting case, the contribution from the processes with $\ket{Q} = \ket{0}$ should be partially canceled out by the processes with $\ket{Q}$ involving two-spin flip states. However, since the double spin-flip states with $k=0$ now form a continuum in the thermodynamic limit, this not only leads to the cancellation of the $\ket{Q} = \ket{0}$ processes, but also results in a tail of signals in the frequency domain \cite{Nandkishore2021}, as schematically shown in Fig.~\ref{fig:PM}(e). Indeed, in the zoomed in data of Fig.~\ref{fig:PM}(f), both the numerical NR and R signals exhibit a tail in the frequency domain. The superposition of slightly different signal frequencies with opposite signs generates the diverging behavior in the time domain data, in line with the ``long-time divergences" discussed in Ref.~\cite{Fava2023}.

\section{Breaking integrability and Confinement}
\label{sec:breaking_integrability}

Having understood the impact of finite size effects, we now consider the addition of a longitudinal field, which introduces a confining potential for the spinon pairs. This explicitly breaks the integrability of the model, meaning it is no longer exactly solvable via a Jordan-Wigner transformation. However, it is possible to obtain analytical results using the two-kink approximation, derived by projecting the full Hamiltonian into the subspace of two-spinon states~\cite{Coldea2010, Rutkevich2008, Rutkevich2010} (see Appendix \ref{sec:twokink} for details of the analytical calculation). In Ref.~\cite{Sim2023}, they explored the impact of confinement on the second-order response $\chi^{(2)}$. Here, we focus on $\chi^{(3)}$, and in particular the evolution of the spinon-echo rephasing signal, predicted as a key fingerprint of fractionalization \cite{Wan2019}, with increasing longitudinal field. 

Figure~\ref{fig:zero_temp_lfield} compares the analytical and ED results for $\chi_{xx}^{(1)}$, $\chi_{xxx}^{(2)}$, and $\chi_{xxxx}^{(3;1,2)}$ for both the weakly confined, with $h_z=0.06$, and strongly confined, with $h_z=0.4$, cases. The confining potential leads to a set of spinon bound states with a discrete energy spectrum $E_n$, clearly visible in the linear response susceptibility $\chi^{(1)}_{xx}$. The energy spacing between these states increases as $h_z$ increases, with for example only the lowest three visible within the frequency range $0 < \omega_t < 6$ in the strongly confined limit, $h_z=0.4$. The second-order susceptibilities are consistent with Ref.~\cite{Sim2023} and are discussed in depth there.  

In the weakly confined limit, $h_z = 0.06$, we see that even a small longitudinal field produces significant deviations from the integrable (zero confinement) results.
First, in $\chi_{xxxx}^{(3;1)}$, there are a sequence of additional peaks surrounding the on-axis pump-probe signal. More interestingly, in $\chi_{xxxx}^{(3;2)}$, the spinon-echo R signal now consists of a grid of discrete peaks spread over a wide frequency range $(2 \leq \omega_t, -\omega_\tau \leq 4)$, in contrast to the sharp anti-diagonal line in the integrable case. The nature of these peaks can be discerned by examining in more detail the structure of the results from the two-kink approximation. The matrix elements involved can be written as $m^x_{0R}m^x_{RQ} m^x_{QP} m^x_{P0}$, with the grid of rephasing peaks corresponding to the case where the ground state is the intermediate state $\ket{Q}=\ket{0}$, and $\ket{P}$ and $\ket{R}$ are spinon bound states (see Fig.~\ref{fig:2kink} in Appendix \ref{sec:twokink} for a plot of the contribution of just the $\ket{Q}=\ket{0}$ case). Thus, we observe peaks at energies $(\omega_t, \omega_\tau) = (E_n, -E_l)$, with intensities proportional to $\abs{m^x_{n0}}^2 \abs{m^x_{l0}}^2$ (peaks on the anti-diagonal have $n=l$, and on the off-diagonal $n\neq l$). These peaks contain similar information to linear response, as it is matrix elements of the form $\abs{m_{n0}}^2$ that appear there as well.     

In the strongly confined limit, $h_z=0.4$, we see that the number of peaks visible has sharply reduced, primarily due to the increased spacing in energy between bound states. In $\chi_{xxxx}^{(3;2)}$, the peaks along the anti-diagonal, as well as their associated cross-peaks, again originate from the case where the ground state is the intermediate state $\ket{Q}=\ket{0}$, and $\ket{P}$ and $\ket{R}$ are spinon bound states. On the other hand, the terahertz rectification (TR) peaks on, and just off, the $\omega_\tau$ axis originate from the case where the intermediate state is a two-kink state (see Fig.~\ref{fig:2kink} in Appendix \ref{sec:twokink} for a plot of just the intermediate two-kink contributions).  

It's important to note that here the discreteness of the spectrum is not a finite-size effect, as it was in the integrable FM phase, but rather it has a clear physical origin. Therefore, the NR signals observed in $\chi_{xxxx}^{(3;1,2)}$ are expected to remain even in the thermodynamic limit. Indeed, the analytical results, qualitatively consistent with the ED data, are computed in the thermodynamic limit (except for system size-dependent intensity factors). Thus, in this particular case, the appearance of a finite NR signal is a direct consequence of confinement converting the spinon continuum into a discrete set of spinon bound states. 

In Fig.~\ref{fig:lfield_detal}, we show the full evolution of $\chi_{xxxx}^{(3;2)}$ with increasing longitudinal field, computed both within the two-kink approximation and ED. The sharp line features of the zero field case, corresponding to free, deconfined spinon pairs, quickly decompose into a discrete set of peaks due to the confining potential. The rephasing signal decomposes into a grid pattern of discrete peaks, a non-rephasing signal appears with a similar grid of diagonal and cross peaks, and the terahertz rectification signal along the $\omega_\tau$ axis also decomposes into discrete peaks which expand along both directions in frequency space. Also note that the two-kink approximation naturally becomes more accurate the larger the longitudinal field. 

Comparing the results from the two-kink approximation and ED, though they agree on most of the important qualitative features, there are some differences. This is due to the fact that we omit four-kink states, and higher, in the two-kink approximation, while there is no such approximation in the ED. For example, in $\chi_{xxxx}^{(3;1)}$, four-kink processes generate a signal at $(\omega_t, \omega_\tau) \approx (E_n, 2E_n)$, which is absent in the two-kink approximation results but which we observe in the ED data. Analogous to the PM phase, the four-kink states also introduce cancellation terms in $\chi_{xxxx}^{(3;1,2)}$, regulating the apparent system size-dependent behavior of R/NR and PP signals from ground state processes, leading ultimately to $L$-independent signals. 

There are also quantitative differences between the two-kink approximation and ED results, especially in $\chi_{xxxx}^{(3;1,2)}$. The analytical predictions tend to overestimate the intensities when compared to the ED data. This is evident in, for example, the pump-probe signal in $\chi_{xxxx}^{(3;1)}$, and the R and NR signals in $\chi_{xxxx}^{(3;1,2)}$. As in the PM phase discussed previously, these differences stem from the proximity of poles of the processes where $\ket{Q}$ is a four-kink state and $\ket{Q}$ is the ground state. In ED, this proximity, coupled with a limited frequency resolution, leads to an intensity reduction due to the cancellation of poles from the two distinct processes. Indeed, time domain data (not shown) reveals an increasing trend in the signal for $\chi_{xxxx}^{(3;1,2)}$ as time $t$ increases, suggesting the presence of two peaks with opposite signs and slightly different frequencies in these signals. 

Finally, it's worth noting that although the two-kink approximation and ED exhibit broadly qualitative agreement for $\chi_{xxxx}^{(3;2)}$, the individual $R_a$ that contribute to it, see Eqn.~(\ref{eq:R}) (and Appendix \ref{sec:2dcs}), are actually quite different in the two methods. For example, the rephasing signal in the two-kink approximation solely comes from $R_2$, while in the ED it comes from a complex partial cancellation of $R_1$ and $R_2$. In some cases, it may thus be useful to not only compare the non-linear susceptibilites, but also the individual $R_a$, to better compare and contrast different approaches. See Appendix \ref{sec:Ra} for more details on this point. 

\begin{figure}[tb]
    \centering
    \includegraphics[width=0.9\columnwidth]{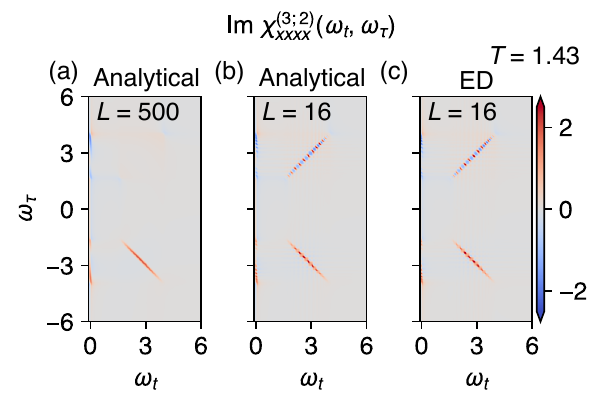}
    \caption{{\bf Finite temperature spectra of 1D-TFIM}. A comparison between analytical (a,b) and numerical (c) results of finite temperature with $L=500$ and $L=16$ is displayed. Parameters $J = 0.7, h_x = 0.3$ are used for calculation.}
    \label{fig:finite_temp}
\end{figure}

\begin{figure}[tb]
    \centering
    \includegraphics[width=\columnwidth]{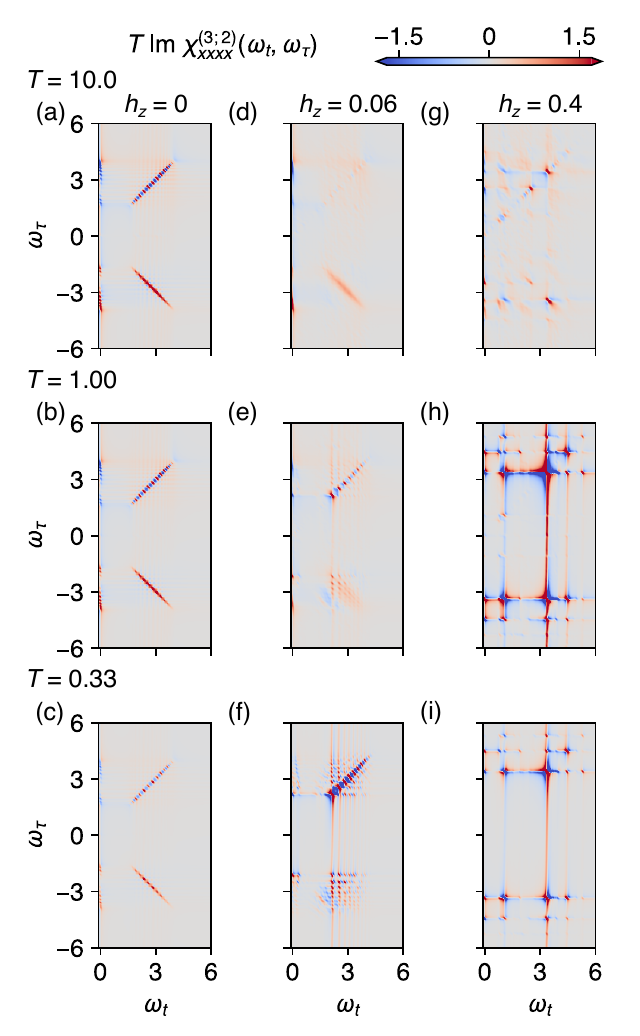}
    \caption{{\bf Temperature dependence of $T\chi^{(3;2)}_{xxxx}(\omega_t, \omega_\tau)$} of the 1D-TFIM in a longitudinal field from ED. The parameters $J = 0.7, h_x = 0.3$. (a-c) $h_z = 0$, i.e. exactly solvable case with inverse temperature $T = 10.0, 1.00, 0.33$, respectively. (d-f) $h_z = 0.06$. (g-i) $h_z = 0.4$.}
    \label{fig:finite_temp_lfield}
\end{figure}


\section{Finite temperature}
\label{sec:finite_temp}

Finally, we make use of one of the other advantages of ED and extend our study to finite temperatures. As in the zero-temperature case, we first investigate the exactly solvable case with $h_z = 0$. Figure~\ref{fig:finite_temp} shows $\chi^{(3;2)}_{xxxx}$ calculated for $L=500$ and $L=16$ at a temperature $T = 1.43$. For this model, one should note that this particular response function is rather special as the only change with temperature is in the intensities of the signals. For $L=500$, the overall peak intensity is suppressed as temperature increases, but the rephasing R signal remains sharp in the two-dimensional frequency space. For $L = 16$, there are twice as many peaks in the signal, which is due to the fact that at finite temperatures we also have a contribution from the parity odd sector of the Jordan-Wigner transformed fermionic Hamiltonian of the TFIM model (see Appendix \ref{sec:tfim} for details). Similar to the zero-temperature case, we also observe a non-rephasing NR signal due to the finite system size, and no other additional signal appears. Figure~\ref{fig:finite_temp}~(c) shows ED result for $L = 16$ at $T = 1.43$. The quantitative agreement between the analytical and ED results helps to validate our method for finite temperature.

Next, we investigate the finite-temperature 2DCS in the presence of the spinon confining potential. The impact on $\chi^{(3;2)}$ is shown in Fig.~\ref{fig:finite_temp_lfield} for the zero, weakly, and strongly confined cases, with the data normalized by $\beta = 1/T$ so that they can be plotted with a common color scale. At a sufficiently low temperature of $T=0.33$, which is below the specific heat peaks for all three cases (see Fig.~\ref{fig:cv} in Appendix \ref{sec:thermal_state}), the intensities and peak positions are close to the zero-temperature result. As temperature increases, the signal intensities decrease for all values of $h_z$. For the weakly confined case, $h_z = 0.06$, the discreteness of the low-temperature signal is smeared out as temperature increases, and the signal becomes continuous. In particular, at $T = 10.0$, the spectrum exhibits qualitative similarities to the $h_z = 0.00$ spectrum found in the thermodynamic limit, suggesting the impact of confinement can only be resolved at low temperatures. For $h_z = 0.4$, the discreteness of the signal is preserved even at high temperatures. We also observe additional peaks along the line $\omega_t \approx 1$, which arise from the processes involving transitions between two-kink states.


\section{Discussion}
\label{sec:discussion}
The calculation of non-linear dynamical response functions for a quantum many-body system is a formidable and challenging task. We explored here, on the one hand, the utility of using ED to calculate non-linear susceptibilities in the 1D-TFIM, and, on the other hand, the impact of confinement on the unique signatures of spinon fractionalization in the model. One of the key advantages of ED, the ability to simulate to long times without any corresponding increase in computational complexity, can be used to obtain high-resolution 2DCS spectra in frequency space. However, the key disadvantage, the relatively small system sizes available with ED, generates additional issues with interpreting the spectra due to potential finite size effects. We have discussed in detail these issues in the context of the 1D-TFIM and its two distinctive ground states. In both limits, ED can capture many of the important qualitative features expected, though care must be taken in distinguishing whether discrete excitations arise due to finite-size effects or due to some underlying physical mechanism. Using this knowledge, we were able to explore the impact of spinon confinement on the third-order susceptibilities, and, in particular, on the characteristic spinon-echo rephasing signal. Combining ED with an analytical two-kink approximation, we were able to show how even a moderate longitudinal field can break up the sharp anti-diagonal signal and induce a visible non-rephasing signal.      

In terms of experimental relevance, there are a number of materials whose dominant interactions can be written in the form of a 1D-TFIM. As an example, in \ch{CoNb_2O_6}, the FM chain of \ch{Co^2+} ions is considered to be a good realization of the 1D-TFIM \cite{Rutkevich2010, Coldea2010, Fava2020, Morris2021}. In this material, the weak interchain interaction effectively introduces a longitudinal field, and the discrete spectrum of spinon bound states has been experimentally observed with linear response probes \cite{Coldea2010, Rutkevich2010}. Therefore, with 2DCS, we would expect qualitative features of the 2D spectrum computed here to also be observed in the material. However, it would be necessary to investigate the effects of additional terms, which are proposed to be relevant in the material, such as XY and bond-dependent interactions, in order to gain a quantitative understanding of the expected experimental spectrum.

\begin{figure}
    \centering
    \includegraphics[width=\columnwidth]{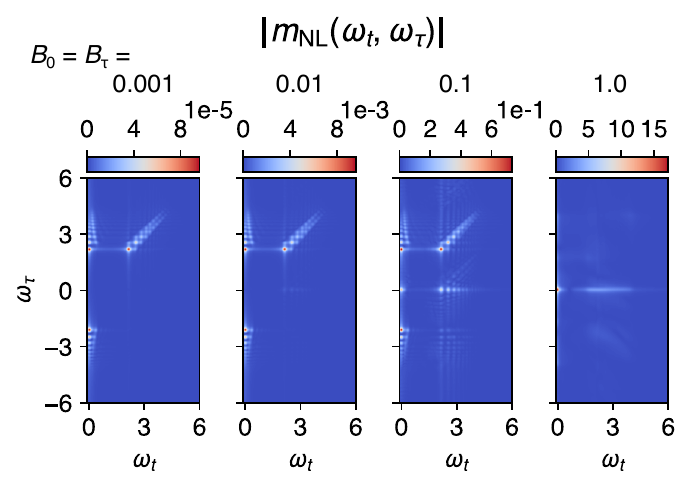}
    \caption{{\bf Fourier transform of the non-linear magnetization} $m_\text{NL}$ with increasing magnitude of incident THz pulses, $B_0,B_\tau$, obtained by ED. The parameters for the weakly confined case, $J = 0.7, h_x = 0.3, h_z = 0.06$ are used.}
    \label{fig:nlm}
\end{figure}

An important caveat, relevant for any experimental comparison, is that in reality it is the non-linear magnetization $m^\alpha_{\text{NL}}$ that is measured in experiment, as opposed to the individual susceptibilities. The non-linear magnetization contains contributions from $\chi^{(2)}$, $\chi^{(3;1)}$, and $\chi^{(3;2)}$, as well as higher order terms. The ratio of the intensities of these terms is, roughly speaking, determined by the strength of the THz field; $n$-th order terms are proportional to the $n$-th power of the THz field. Therefore, if there is a finite $\chi^{(2)}$, it will typically be the dominant contribution to the experimental signal. Figure~\ref{fig:nlm} shows the Fourier transform of the non-linear magnetization $m^x_\text{NL}$  obtained by ED with different strengths of the magnetic field pulses. The parameters for the weakly confined case, $J = 0.7, h_x = 0.3, h_z = 0.06$ are used. Up to $B_0 = B_\tau = 0.01$, the visible signal is dominated by the $\chi^{(2)}$ contribution. Only when $B_0 = B_\tau = 0.1$ is the visibility of the third-order contribution comparable to the second-order one, including the third-order pump-probe and rephasing signals. Finally, at $B_0 = B_\tau = 1.0$, the field pulses can no longer really be considered as perturbations, and the signal becomes dominated by the pump-probe response. 

Looking forward, there are a multitude of non-linear many-body phenomena that can be further studied using ED and complementary analytical insights. For example, in ordered phases, the 2DCS signatures of non-linear magnon decay and lifetime effects, or the interplay between distinct excitation modes. On the computational front, it would be useful to test the capabilities of ED in studying 2D spin models. As an example, there are already known analytical results for the non-linear response of the Kitaev honeycomb model \cite{Choi2020,Qiang2023,Kazem2023}, which can be used to compare and contrast with ED results and to diagnose potential finite size effects. Taken together, it's clear that there is a wide world of exciting physics beyond the linear response regime still waiting to be explored.

\acknowledgments
We acknowledge partial funding from the DFG within Project-ID 277146847, SFB 1238 (projects C02, C03). 
This work has benefitted from multiple exchanges during the 2023 KITP workshop ``A New Spin on Quantum Magnets", 
supported in part by grant NSF PHY-1748958 to the Kavli Institute for Theoretical Physics (KITP).
The numerical simulations were performed on the JUWELS cluster at the Forschungszentrum Juelich 
and the Noctua2 cluster at PC2 in Paderborn.

\newpage
\appendix

\section{Full expressions for the non-linear susceptibilities}
\label{sec:2dcs}

We complement our technical discussion in the main manuscript here by providing the full expressions for the non-linear susceptibilities, $\chi^{(2)}$ and $\chi^{(3)}$. The second-order susceptibility can be expressed as
\begin{equation}
       \chi^{(2)}_{\alpha\beta\gamma}(t_2,t_1) = -\frac{2}{N}\text{Re}[ R_1 - R_2],
\end{equation}
%
with
\begin{equation}
    \begin{split}
    R_1 &= \avg{M^\alpha(t_2 + t_1) M^\beta(t_1)M^\gamma(0)},\\
    &= \sum_{PQ} m^\alpha_{0Q}m^\beta_{QP} m^\gamma_{P0} e^{-\text{i}E_P t_1} e^{-\text{i}E_Q t_2},\\
    R_2 &= \avg{M^\beta(t_1) M^\alpha(t_2+t_1)M^\gamma(0)},\\
    &= \sum_{PQ} m^\beta_{0Q}m^\alpha_{QP} m^\gamma_{P0} e^{-\text{i}E_P t_1} e^{-\text{i}(E_P - E_Q) t_2}. 
    \end{split}
\end{equation}

The third-order susceptibility $\chi^{(3)}$ can be expressed as
\begin{equation}
           \chi^{(3)}_{\alpha\beta\gamma\delta}(t_3,t_2,t_1) = \frac{2}{N}\text{Im}[ R_1 + R_2 + R_3 + R_4],
\end{equation}
with
\begin{equation}
    \begin{split}
    R_1 &= \avg{M^\gamma(t_1) M^\beta(t_2+t_1)M^\alpha(t_3+t_2+t_1) M^\delta(0)}\\
    &= \sum_{PQR} m^\gamma_{0R}m^\beta_{RQ} m^\alpha_{QP} m^\delta_{P0} \\
    &\ \ \ \ \times e^{-\text{i}E_P t_1} e^{-\text{i}(E_P - E_R) t_2} e^{-\text{i}(E_P - E_Q)t_3},\\
    R_2 &= \avg{M^\delta(0) M^\beta(t_2+t_1)M^\alpha(t_3+t_2+t_1) M^\gamma(t_1)}\\
    &= \sum_{PQR} m^\delta_{0R}m^\beta_{RQ} m^\alpha_{QP} m^\gamma_{P0} \\
    &\ \ \ \ \times e^{\text{i}E_R t_1} e^{-\text{i}(E_P - E_R) t_2} e^{-\text{i}(E_P - E_Q)t_3}\\
    R_3 &= \avg{M^\delta(0) M^\gamma(t_1)M^\alpha(t_3+t_2+t_1) M^\beta(t_2+t_1)}\\
    &= \sum_{PQR} m^\delta_{0R}m^\gamma_{RQ} m^\alpha_{QP} m^\beta_{P0} \\
    &\ \ \ \ \times e^{\text{i}E_R t_1} e^{\text{i}E_Q t_2} e^{-\text{i}(E_P - E_Q)t_3},\\
    R_4 &= \avg{M^\alpha(t_3+t_2+t_1) M^\beta(t_2+t_1)M^\gamma(t_1) M^\delta(0)}\\
    &= \sum_{PQR} m^\alpha_{0R}m^\beta_{RQ} m^\gamma_{QP} m^\delta_{P0} \\
    &\ \ \ \ \times e^{-\text{i}E_P t_1} e^{-\text{i}E_Q t_2} e^{-\text{i}E_Rt_3}. 
    \end{split}
    \label{eq:Rs}
\end{equation}

\section{Preparation of the thermal state}
\label{sec:thermal_state}
The thermal state $\left|\phi\right>$, which is used to calculate specific heat and 2DCS, is obtained by the imaginary time evolution of the initial state $\left|\phi_0\right>$.
Instead of computing $\left|\phi\right>$ directly, we first compute $e^{\beta(l - H)/2}\ket{\phi_0}$, where $l$ is set to larger than the largest eigenvalue of $H$.
The expansion of the exponential operator is given by:
\begin{equation}
    \begin{split}
        e^{\beta(l - H)/2}\ket{\phi_0} & = \sum_{k=0}^\infty \frac{(\beta/2)^k}{k!}(l - H)^k\ket{\phi_0}\\
        &= \sum_{k=0}^\infty \frac{(\beta/2)^k}{k!}\ket{k},
    \end{split}
\end{equation}
It is known that the relevant terms in the sum is localized in the range $\abs{k^* - k} = o(L)$, where the temperature of normalized microcanonical thermal pure quantum state $\ket{\phi_{k^*}} = \ket{k^*}/\Vert \ket{k^*}\Vert$  is close to $1/\beta$ \cite{Sugiura2012, Sugiura2013}. We choose the number of iterations as $k_{\text{max}} = 2000$, which is sufficiently large for the temperature range we are interested in in this study. Then, multiplying scalar $e^{-\beta l/2}$ to this state gives $\left|\phi\right>$. Similarly to the linear response \cite{Iitaka2003}, we calculate the non-linear susceptibility from the time evolution of $\ket{\phi}$.

\begin{figure}[tb]
    \centering
    \includegraphics[width=0.8\columnwidth]{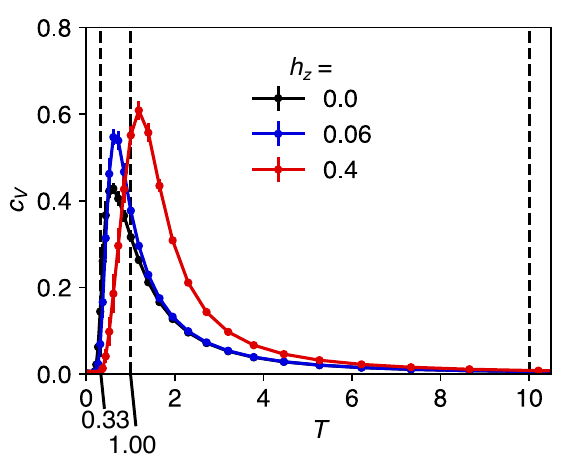}
    \caption{{\bf Specific heat $c_V$ of 1D-TFIM in the longitudinal field} from ED with $L=16$. The parameters $J = 0.7, h_x = 0.3$, and several values of $h_z$ are considered.}
    \label{fig:cv}
\end{figure}

Figure.~\ref{fig:cv} shows the calculated specific heat $c_V$ with different values of $h_z$. The temperatures used for the finite temperature 2DCS are chosen to cover the different temperature regimes separated by the peaks of $c_V$.

\section{Zero/Finite temperature 2DCS in TFIM}
\label{sec:tfim}
Here, we provide a 2DCS in TFIM based on Ref.~\cite{Wan2019}. With the Jordan-Wigner transformation,
\begin{equation}
    \begin{split}
        \sigma^x_i &= 2c_i^\dagger c_i - 1,\\
        \sigma^z_i &= (c_i^\dagger + c_i)\exp\left(\text{i}\pi\sum_{j=1}^{i-1}c_j^\dagger c_j\right),
    \end{split}
\end{equation}
the Hamiltonian reads
\begin{equation}
    \begin{split}
        \mathcal{H}_p &= - J\sum_{i=1}^{L-1}(c_i^\dagger - c_i)(c_{i+1}^\dagger + c_{i+1}) \\
        +\ & (-1)^p J(c_L^\dagger - c_L)(c_1^\dagger + c_1) 
        - h_x\sum_{i=1}^L(2n_i - 1),
    \end{split}
\end{equation}
where $p$ is the parity of the number of fermions. Among $2^L$ Fock states of each fermionic Hamiltonian, only $2^{L-1}$ states with the same parity as $p$ are correct states of the spin Hamiltonian with the periodic boundary condition. The ground state is known to lie in the parity even sector, therefore the contribution from parity odd sector becomes finite only at the finite temperature 2DCS. We introduce the momentum representation of the fermionic operators as $c_j = (1/\sqrt{L})\sum_k e^{\text{i}kj}c_k$, where $k = \pm(2n - 1)\pi/L$ with $n = 1, \cdots, L/2$ for parity even sector, and $k$ satisfies $k = 2\pi n/L$ with $n = -L/2 + 1, \cdots, 0, \cdots, L/2$ for parity odd sector. Next, we introduce a pair Hamiltonian with momentums $k$ and $-k$ as
\begin{equation}
    \begin{split}
        \mathcal{H}_k &= (c_k^\dagger, c_{-k})
        \left( \begin{array}{cc}
            \epsilon_k & \text{i}\Delta_k \\
            -\text{i}\Delta_k & -\epsilon_k
        \end{array} \right)
        \left( \begin{array}{c}
            c_k \\
            c_{-k}^\dagger
        \end{array} \right),
    \end{split}
\end{equation}
where $\epsilon_k = - 2J\cos k - 2h_x$ and $\Delta_k = -2J\sin k$, and a Hamiltonian for $k = 0$ and $k = \pi$ of the parity odd sector as
\begin{equation}
    \mathcal{H}_{k=0,\,\pi} = -2J(n_0 - n_\pi) - 2h_x(n_0 + n_\pi - 1).
\end{equation}
Now, we can rewrite the Hamiltonian of the even sector as
\begin{equation}
    \mathcal{H}_0 = \sum_{k>0}\mathcal{H}_k,
\end{equation}
and the Hamiltonian of the odd sector as
\begin{equation}
    \mathcal{H}_1 = \sum_{0 < k < \pi}\mathcal{H}_k + \mathcal{H}_{k=0,\,\pi}.
\end{equation}
Performing the Bogoliubov transformation:
\begin{equation}
    \left( \begin{array}{c}
        c_k \\
        c_{-k}^\dagger
    \end{array} \right)
    = \left( \begin{array}{cc}
        \cos\frac{\theta_k}{2} & -\text{i}\sin\frac{\theta_k}{2} \\
        -\text{i}\sin\frac{\theta_k}{2} & \cos\frac{\theta_k}{2}
    \end{array} \right)
    \left( \begin{array}{c}
        \gamma_k \\
        \gamma_{-k}^\dagger
    \end{array} \right),
\end{equation}
where $\tan\theta_k = \Delta_k/\epsilon_k$, we obtain the diagonalized Hamiltonian
\begin{equation}
    \mathcal{H}_k = \lambda_k(\gamma_k^\dagger\gamma_k - \gamma_{-k}\gamma_{-k}^\dagger),
\end{equation}
where $\lambda_k = \sqrt{\epsilon_k^2 + \Delta_k^2}$.

Using the Anderson pseudo-spins:
\begin{equation}
    \begin{split}
        \tau_k^x &= \gamma_{-k}\gamma_k + \gamma_k^\dagger\gamma_{-k}^\dagger, \\
        \tau_k^y &= \text{i}(\gamma_{-k}\gamma_k - \gamma_k^\dagger\gamma_{-k}^\dagger), \\
        \tau_k^z &= \gamma_k^\dagger\gamma_{k} - \gamma_{-k}\gamma_{-k}^\dagger,
    \end{split}
\end{equation}
magnetization $M^x$ is given by
\begin{equation}
    \begin{split}
        M^x &= \frac{1}{2}\sum_{i=1}^L\sigma_i^x \\
        &= \sum_{0<k<\pi}\left(\cos\theta_k\tau_k^z + \sin\theta_k\tau_k^y\right),\\
        &\equiv \sum_{0<k<\pi}m_k^x,
    \end{split}
\end{equation}
Note that we drop the contribution from $k = 0$ and $k = \pi$ in the odd sector, which is a constant of the motion and does not contribute to the 2DCS.
In the Heisenberg picture, $m_k^x$ becomes:
\begin{equation}
    m_k^x(t) = \cos\theta_k\tau_k^z + \sin\theta_k(\tau_k^y \cos(2\lambda_k t) + \tau_k^x \sin(2\lambda_k t)).
\end{equation}

Using the fact that each $m_k^x$ commutes with each other, the third-order response is given by
\begin{widetext}
    \begin{equation}
        \begin{split}
            \chi^{(3)}_{xxxx}(t_3, t_2, t_1) 
            &= \frac{1}{L}\sum_{0<k<\pi}\left<[[[m_k^x(t_3+t_2+t_1), m_k^x(t_2 + t_1)], m_k^x(t_1)], m_k^x(0)]\right> \\
            &= \frac{1}{L}\sum_{0<k<\pi} \left[A_k^{(1)} + A_k^{(2)} + A_k^{(3)} + A_k^{(4)}\right],\\
        \end{split}
    \end{equation}
\end{widetext}
where
\begin{equation}
    \begin{split}
        A_k^{(1)} &= 8\sin^2\theta_k\cos^2\theta_k\sin(2\lambda_k (t_3 + t_2 + t_1))\left<\tau_k^z\right>, \\
        A_k^{(2)} &= -8\sin^2\theta_k\cos^2\theta_k\sin(2\lambda_k (t_2 + t_1))\left<\tau_k^z\right>, \\
        A_k^{(3)} &= 4\sin^4\theta_k\sin(2\lambda_k (t_3 + t_1))\left<\tau_k^z\right>, \\
        A_k^{(4)} &= 4\sin^4\theta_k\sin(2\lambda_k (t_3 - t_1))\left<\tau_k^z\right>, 
    \end{split}
\end{equation}
Here we utilized the fact that $\left<\tau_k^x\right> = \left<\tau_k^y\right> = 0$.
Zero temperature limit Eq.~(\ref{eq:chi3_tfim}) is obtained by $\left<\tau_k^z\right> = -1$, and consider the even sector only, i.e., we take a sum for $k = (2n - 1)\pi/L$ with $n = 1, \cdots, L/2$.

Finite temperature 2DCS is obtained by taking the thermal average of $\left<\tau_k^z\right>$. The distribution function is given by
\begin{equation}
    Z = \text{Tr}\left(e^{-\beta\mathcal{H}_\text{spin}}\right) = \sum_p\text{Tr}\left(P_pe^{-\beta\mathcal{H}_p}\right) \equiv \sum_p Z_p,
\end{equation}
where $P_p$ is the projection operator to the parity $p$ sector defined as
\begin{equation}
    P_p = \frac{1}{2}\left(1 + (-1)^p e^{-\text{i}\pi N}\right).
\end{equation}
$Z_0$ can be computed as:
\begin{equation}
    Z_0 = \frac{e^{-\beta E_0}}{2}\left[\prod_k(1 + e^{-\beta\lambda_k}) + \prod_k(1 - e^{-\beta\lambda_k})\right],
\end{equation}
where $E_0 = -\sum_{k>0}\lambda_k$ is the ground state energy of the parity even sector, and the product over $k$ runs the entire Brillouin zone.
Similarly, $Z_1$ can be computed as:
\begin{equation}
    Z_1 = \frac{e^{-\beta E_1}}{2}\left[\prod_k(1 + e^{-\beta\lambda_k}) - \prod_k(1 - e^{-\beta\lambda_k})\right],
\end{equation}
Where $E_1 = -\sum_{0 < k < \pi}\lambda_k + 2h_x$ is the energy of the state obtained by annihilating the $k=0$ fermion from the odd-sector ground state. For $k=0$ and $\pi$ we take $\lambda_0 = -2J - 2h_x$ and $\lambda_\pi = 2J - 2h_x$, respectively. Finally, we obtain
\begin{equation}
    \begin{split}
        &\left<\tau_{k}^z\right> = -\frac{
            Z_{p;\bar{k}}\left(1 - e^{-2\beta\lambda_k}\right)
            }{Z_0 + Z_1}
    \end{split}
\end{equation}
where $p$ is the parity that is consistent with $k$, and
$Z_{p;\bar{k}}$ is obtained by omitting the contribution from $\pm k$ in the product over $k$.

\section{Phase untwisted 2DCS}
\label{sec:posneg_tau}

\begin{figure}[b]
    \centering
    \includegraphics[width=\columnwidth]{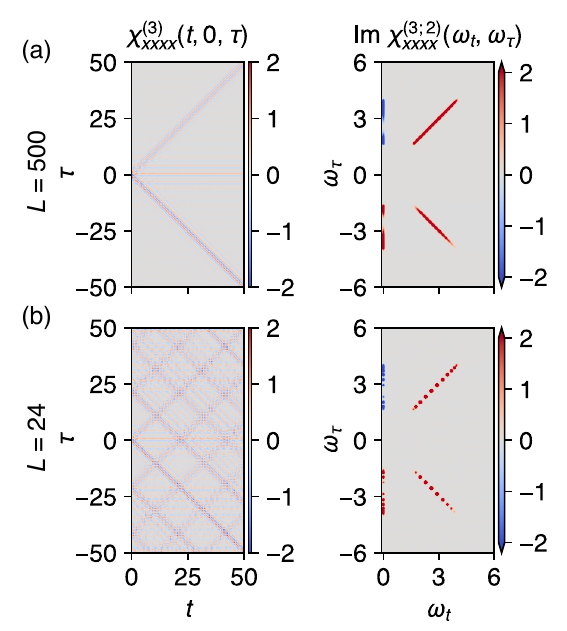}
    \caption{{\bf Phase untwisted 2DCS} obtained by including negative $\tau$ in the calculation. The parameters $J = 0.7, h_x = 0.3, h_z = 0.06$, and the system size $L =500$ (a) and $L = 24$ (b) are used.}
    \label{fig:posneg_tau}
\end{figure}

In the main text, we discussed how the positive time constraint leads to a distorted signal known as phase twisting, which can even eliminate the NR signal at $\omega_t = \omega_\tau$. We demonstrate that this issue can be resolved by incorporating negative $\tau$ values, which is feasible in analytical calculations and ED simulations. Figure~\ref{fig:posneg_tau} presents the 2DCS results obtained by including negative $\tau$. It is noteworthy that even for a substantially large system size ($L=500$), the NR signal is observable. Furthermore, the data for $L=24$ now appears as a discretized version of the $L=500$ data. It is important to note that NR and R signals are not symmetric, as predicted by Eq.~\ref{eq:chi3_tfim}. Therefore, phase untwisting cannot be simply achieved by symmetrizing the experimental data around about $\omega_\tau$ \cite{Hart2023}.

\section{Two-kink model}\label{sec:twokink}

Here, we provide the detailed calculation of 2DCS within the TK model following the discussion in Refs.~\cite{Rutkevich2008, Rutkevich2010, Sim2023}. $\ket{j,l} \equiv \ket{\cdots\uparrow\uparrow\downarrow_j\cdots\downarrow_{(j+l-1)}\uparrow\uparrow\cdots}$ is the state with two kinks at $j$ and $j+l-1$. The projection to the subspace spanned by $\ket{j,l}$ gives TK Hamiltonian which acts as:
\begin{equation}
    \begin{split}
    \mathcal{H}_{\text{TK}}\ket{j,l} & = (4J + 2h_z)\ket{j,l}  \\
    & - h_x\left(\ket{j,l+1} + \ket{j, l-1}\right. \\
    & + \left.\ket{j+1, l-1} + \ket{j-1, l+1}\right).
    \end{split}
\end{equation}
Next, we introduce the momentum basis $\ket{p, l} = (1/\sqrt{L})\sum_j e^{\text{i}pj}\ket{j,l}$.
Here, normalization factor $1/\sqrt{L}$ is introduced to make the qualitative comparison with ED possible. As far as the zero-temperature 2DCS is concerned, we can restrict ourselves to the zero-momentum sector. Using $\ket{l} \equiv \ket{p = 0, l}$, we can write the TK Hamiltonian as
\begin{equation}
    \begin{split}
    \mathcal{H}_{\text{TK}}\ket{l} & = (4J + 2h_z)\ket{l}  \\
    & - 2h_x(\ket{l+1} + \ket{l-1}).
    \end{split}
\end{equation}
Introducing the $n$-th eigen state $\ket{\Phi_n} = \sum_l \psi_n(l)\ket{l}/\sqrt{\sum_{l=1}^\infty\abs{\psi_n(l)}^2}$ and its eigen energy $E_n$ the Schr\"odinger equation becomes
\begin{equation}
    \mathcal{H}_{\text{TK}}\ket{\Psi_n} = E_n\ket{\Psi_n}.
\end{equation}
Using dimensionless paramters $\lambda_n = (E_n - 4J)/(4h_x)$ and $\mu = h_z/(2h_x)$, we obtain the following recursion relation:
\begin{equation}
    \begin{split}
     & (- \lambda_n + \mu l)\psi_n(l) \\
     &- (\psi_n(l+1) + \psi_n(l-1))/2 = 0,
    \end{split}
\end{equation}
with the boundary conditions $\psi_n(0) = \psi_n(+\infty) = 0$.
Eigen values $\lambda_n$ can be obtained using the zeros $\nu_n$ of the Bessel function $J_\nu(1/\mu)$ as a function of its order $\nu$ as
\begin{equation}
    \lambda_n = -\mu\nu_n,
\end{equation}

Now, we consider processes where each intermediate state is a two-kink state. For example, $\chi^{(3)}_{xxxx}$ is given by:
\begin{equation}
    \begin{split}
    &\chi^{(3)}_{xxxx}(t_3, t_2, t_1)|_{\ket{Q} = \ket{\Phi_n}} \\
    &= \frac{2}{L}\sum_{nml}\sum_{p=1}^{4}\text{Im}\Big[\braket{0|M^x|\Phi_n}\braket{\Phi_n|M^x|\Phi_m}\\
    &\times\braket{\Phi_m|M^x|\Phi_l}\braket{\Phi_l|M^x|0} C_{p}(t_3,t_2,t_1)\Big]\\
    &\equiv \sum_{nml}\sum_{p=1}^{4} \text{Im}\Big[A_{n,m,l} C_{p}(t_3,t_2,t_1)\Big]
    \end{split}
\end{equation}
where $R_p$ defined in Eq.~(\ref{eq:Rs}) is decomposed into the matrix elements $A_{n,m,l}$ and the time-dependent part $C_p(t_3, t_2, t_1)$. The factor $(2/L)$ is incorporated into $A_{n,m,l}$. Here, the ferromagnetic ground state $\ket{0}$ is defined as $\prod_{j=1}^L\ket{\uparrow}_j$, with an energy $E_0 = 0$. Following the discussion outlined in Ref.~\cite{Sim2023}, we obtain

\begin{equation}
    \braket{\Phi_n|M^x|0} = \frac{\sqrt{L}\psi_n(1)}{\sqrt{\sum_{l=1}^\infty\abs{\psi_n(l)}^2}}
    \label{eq:zero_Mx_n}
\end{equation}
and 
\begin{equation}
    \begin{split}
    &\braket{\Phi_n|M^x|\Phi_m} \\
    &= -\frac{\psi_n(1)\psi_m(1)}{\sqrt{\sum_{l=1}^\infty\abs{\psi_n(l)}^2}\sqrt{\sum_{l=1}^\infty\abs{\psi_m(l)}^2}}\\
    & \times \left(\frac{1}{\lambda_n - \lambda_m - \mu} + \frac{1}{\lambda_m - \lambda_n - \mu}\right).
    \end{split}
\end{equation}
Therefore $A_{n,m,l}$ are obtained as
\begin{equation}
    \begin{split}
        & A_{n,m,l}
        = 2\left(\frac{1}{\lambda_n - \lambda_m - \mu} + \frac{1}{\lambda_m - \lambda_n - \mu}\right)\\
        & \times \left(\frac{1}{\lambda_m - \lambda_l - \mu} + \frac{1}{\lambda_l - \lambda_m - \mu}\right)I_nI_mI_l
    \end{split}
\end{equation}
where the relative intensity of the $n$-th mode is calculated as:
\begin{equation}
    \begin{split}
        I_n & = \frac{\abs{\psi_n(1)}^2}{\sum_{l=1}^\infty\abs{\psi_n(l)}^2}\\
         & = 2\mu\left.\left\{\frac{\partial}{\partial\nu}\left[\frac{J_\nu(1/\mu)}{J_{\nu+1}(1/\mu)}\right]\right\}^{-1}\right|_{\nu \rightarrow \nu_n},
    \end{split}
\end{equation}
which does not depend on the specific choice of $\psi_n(1)$. Note that $\chi^{(3)}_{xxxx}|_{\ket{Q} = \ket{\Phi_n}}$ does not depend on $L$.

\begin{figure}
    \centering
    \includegraphics[width=0.9\columnwidth]{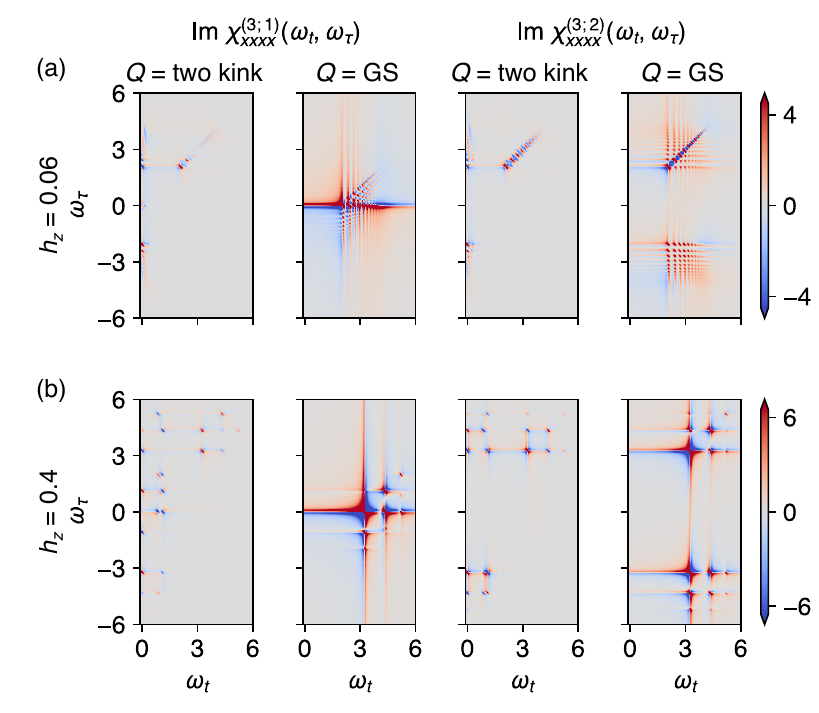}
    \caption{Separate calculation of the contributions from the processes involving $\ket{Q} = \ket{\Phi_n}$, i.e., two-kink states, and $\ket{Q} = \ket{0}$ to $\chi^{(3;1,2)}_{xxxx}$ within two-kink approximation. The parameters $J = 0.7, h_x = 0.3, h_z = 0.06$ (a) and $h_z = 0.4$ (b) are used. }
    \label{fig:2kink}
\end{figure}

As can be seen in Fig.~\ref{fig:2kink}, the signal obtained by only considering two-kink states as intermediate states does not reproduce the many of features obtained by ED. In particular, the PP in $\chi^{(3;1)}_{xxxx}$ and R in $\chi^{(3;2)}_{xxxx}$ are entirely missing.  To obtain a non-rephasing signal, we further consider the process that involves $\ket{Q} = \ket{0}$. Even though, as discussed before, the contribution from such process is expected to be mostly canceled out by the processes that involve four-kink states, we found that the inclusion of this process gives a better agreement about the location of the peaks with ED data. 

For example, using Eq.~(\ref{eq:zero_Mx_n}), $\chi^{(3)}_{xxxx}(t, 0, \tau)|_{\ket{Q} = \ket{0}}$ can be obtained as
\begin{equation}
    \begin{split}
        &\chi^{(3)}_{xxxx}(t, 0, \tau)|_{\ket{Q} = \ket{0}} =
        2L\sum_{n,l}I_nI_l \\
        &\times \text{Im}\left[e^{-\text{i}E_n t} e^{-\text{i}E_n\tau} + 2e^{\text{i}E_l\tau} e^{-\text{i}E_n t} + e^{-\text{i}E_n \tau} e^{-\text{i}E_l t}\right]
    \end{split}
\end{equation}
The second term of the left-hand side gives the signals at the fourth quadrant, i.e. $(\omega_t, \omega_\tau) = (E_n, -E_l)$ with intensity $2LI_nI_l$ (Fig.~\ref{fig:2kink}), proportional to $L$. This extensive terms should be canceled out by the processes that involve four-kink states, making the remaining signal $L$ independent. 

\section{Separate calculation of each $R_a$}\label{sec:Ra}

\begin{figure}
    \centering
    \includegraphics[width=0.9\columnwidth]{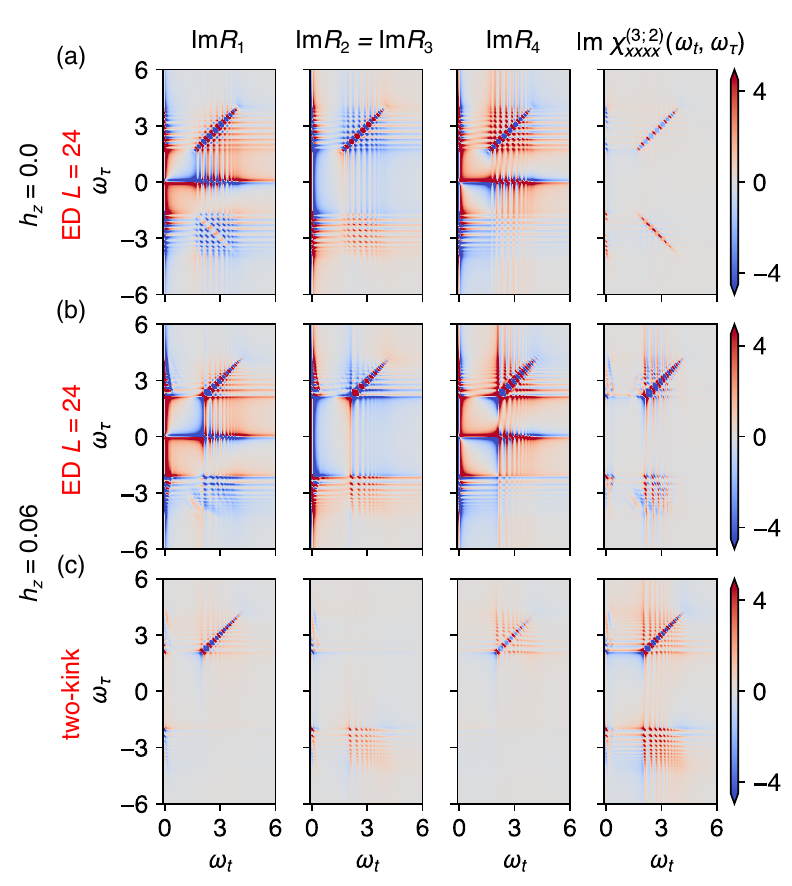}
    \caption{Separate calculation of each $R_{p}$ using ED (a, b) and two-kink model (c). When $t_1 = \tau, t_2 = 0, t_3 = t$, the relation $R_2 = R_3$ holds. Parameters $J = 0.7, h_x = 0.3$ are used.}
    \label{fig:rs}
\end{figure}

In the main text, we numerically apply a magnetic field in a manner similar to the MPS study in Ref.~\cite{Gao2023}. Alternatively, we can directly calculate the imaginary part of $R_a$ using ED, which is more similar to the technique used in Ref.~\cite{Sim2023}. Even though each $R_a$ is not accessible by experiment, it can be useful to understand which processes contribute to each signal. Here we provide an example in Fig.~\ref{fig:rs} where we calculate the imaginary part of each $R_a$ using ED and the two-kink approximation. 

Figure~\ref{fig:rs}(a) shows the case of $h_z = 0.0$, i.e. the exactly solvable case. In addition to the peaks in $\chi^{(3;2)}_{xxxx}$, we observed additional signals at $\omega_t = 0$ and $\omega_\tau = 0$. These contributions come from the process involving the transition from $\ket{0}$ to $\ket{0}$ under the action of $M_x$. $m^{x}_{00}$ is finite because of the finite value of $h_x$. These contributions cancel each other out, giving zero intensity in $\chi^{(3;2)}_{xxxx}$. Furthermore, in addition to the anti-diagonal rephasing signal, we found off-diagonal rephasing signals of opposite sign in $R_1$ and $R_2$. This can be understood as follows: The rephasing signals in $R_2$ come from the process where $\ket{Q} = \ket{0}$. Similar to what we saw in the two-kink calculation, this results in a grid of peaks at $(\omega_t, \omega_\tau) = (E_{k_1}, -E_{k_2})$, where pairs of spinons with momentum $(k_1, -k_1)$ and $(k_2, -k_2)$ are created in the intermediate stage. On the other hand, the rephasing signals in $R_1$ come from the process involving $\ket{Q} = \ket{k_1, -k_1, k_2, -k_2}$, i.e. four-kink states. Since two pairs of kinks are non-interacting, we can freely create one on top of the other. An exception is when $k_1 = k_2$, in which case the creation of the second pair of kinks is forbidden due to the fermionic nature of the spinons, which explains the absence of peaks at $\omega_t = -\omega_\tau$ in $R_1$. After taking the sum of $R_1 + 2R_2 + R_4$, these off-diagonal parts of the rephasing signals are exactly canceled out, and we obtain only a anti-diagonal rephasing signal, where no cancellation occurs due to the absence of the rephasing signal in $R_1$ at $\omega_t = -\omega_\tau$.

Figure~\ref{fig:rs}(b) shows the case of $h_z = 0.06$. Again we see a rephasing signal in $R_1$ and $R_2$. The interacting nature of the confined spinons makes the positions and intensities of the peaks different from each other, and the resulting $\chi^{(3;2)}_{xxxx}$ signal has finite off-diagonal peaks. In the two-kink approximation [Fig.~\ref{fig:rs}(c)] we do not see any rephasing signals in $R_1$, which is consistent with the fact that we have not included four-kink states in the calculation. We also do not see a rather strong peak at $\omega_t = 0$ and $\omega_\tau = 0$, again consistent with the fact that we did not include the process involving the transition from $\ket{0}$ to $\ket{0}$ under the action of $M_x$.


\bibliography{ref}

\end{document}